\documentclass[twocolumn,superscriptaddress,prl,showpacs]{revtex4}
\usepackage{graphicx}
\usepackage{epsfig}
\usepackage{amsmath, amssymb}
\usepackage{verbatim}
\usepackage{bm}
\usepackage{wasysym}
\usepackage{color}

\newcommand{\gammabar}{\ensuremath\gamma\kern-0.53em-}

\begin{document}

\title{Generalized Kitaev Models and Slave Genons}
\author{Maissam Barkeshli}
\affiliation{Microsoft Station Q, Santa Barbara, CA, 93106, USA}
\author{Hong-Chen Jiang}
\affiliation{Department of Physics, University of California, Berkeley, California 94720, USA}
\affiliation{Stanford Institute for Materials and Energy Sciences, SLAC National Accelerator Laboratory, Menlo Park, CA 94025, USA}
\author{Ronny Thomale}
\affiliation{Institute for Theoretical Physics, University of W\"urzburg, Am Hubland, D-97074 W\"urzburg, Germany}
\author{Xiao-Liang Qi}
\affiliation{Department of Physics, Stanford University, Stanford, CA 94305 }

\begin{abstract}
We present a wide class of partially integrable lattice models with
two-spin interactions, which generalize the Kitaev honeycomb model. These models
have an infinite number of conserved quantities associated with each plaquette
of the lattice, conserved large loop operators on the torus, and protected topological
degeneracy. We introduce a `slave-genon' approach, which generalizes the
Majorana fermion approach in the Kitaev honeycomb model. The Hilbert space of our
spin model can be embedded into an enlarged Hilbert space of
non-Abelian twist defects, referred to as genons. In the enlarged
Hilbert space, the spin model is exactly reformulated as a model of
non-Abelian genons coupled to a discrete gauge field. We discuss in
detail a particular $Z_3$ generalization,
and show that in a certain limit the model is analytically tractable and may produce a non-Abelian topological phase with chiral parafermion edge states. 
\end{abstract}

\pacs{71.10.Pm}

\maketitle

%Supp~\cite{supplement}

\it Introduction--\rm The Kitaev honeycomb model \cite{kitaev2006} is an exactly solvable spin model
on the two-dimensional hexagonal lattice, which can realize different exotic
topologically ordered phases of matter, along with non-Abelian quasiparticle excitations.
Over the past decade, this model has generated remarkable excitement\cite{nussinov2013}:
its solvability has provided a theoretical framework to study the emergence
of topological order and non-Abelian anyons from microscopic models, while its simplicity
% RT: has given rise to the possibility that it may be experimentally realized,
supports the hope for experimental realization,
either in Mott insulators with strong spin orbit coupling, such as various
Iridate compounds \cite{chaloupka2010,singh2012}, or directly engineered
with designer Hamiltonians \cite{duan2003}. In particular, the non-Abelian state in the
Kitaev model would open the possibility of topological quantum computation \cite{nayak2008}.

In this paper, we generalize the Kitaev honeycomb model to a much larger
class of partially integrable spin models with only nearest-neighbor interactions.
We show that there is an exact transformation whereby these models can be reformulated
in terms of an array of interacting non-Abelian defects coupled to a static discrete gauge
field. In order to implement the exact transformation, we introduce a ``slave genon'' approach,
where the local Hilbert space on each site is rewritten in terms of the topological degeneracy of
a set of extrinsic non-Abelian twist defects, referred to as genons \cite{barkeshli2012a,barkeshli2013genon},
together with a constraint on their overall fusion channel. This generalizes the
Majorana fermion representation of the original Kitaev honeycomb model \cite{kitaev2006}.
While the transformed problem is itself a non-trivial interacting problem, certain results in
1+1 dimensional critical phenomena can then be utilized to solve the model in certain limits.

{We will focus on a particular $Z_n$ rotor generalization of the Kitaev model for
most of the paper, and discuss more general models in the end of the draft.} We
introduce a graphical method to perform the slave genon technique,
making use of genons in bilayer FQH states \cite{barkeshli2012a,barkeshli2013genon}, with
a $1/n$ Laughlin state in each layer. In the case $n=2$, the genons localize Majorana fermion
zero modes, thus reproducing Kitaev's construction. More generally they localize parafermion zero
modes \cite{barkeshli2012a,barkeshli2013genon,fendley2012,lindner2012,clarke2013,cheng2012,vaezi2013,
barkeshli2013defect,mong2013}. For the case $n=3$, we present some preliminary
numerical results, and discuss the possible realization of a non-Abelian $Z_3$ parafermion
phase, which contains the non-Abelian Fibonacci anyon \cite{nayak2008} in its excitation spectrum.
\begin{figure}
\centerline{
\includegraphics[width=1.75in]{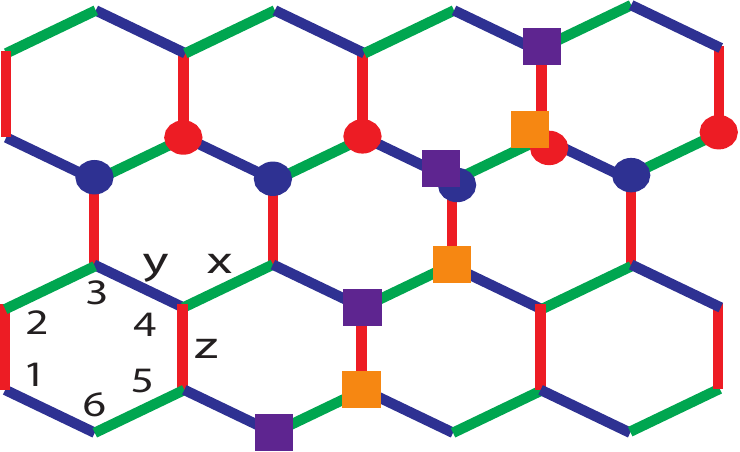}
}
\caption{
\label{fig1}
The links of a honeycomb lattice are labelled x, y, or z, depending on their orientation.
Sites on a plaquette are labelled 1,..,6, as shown. Red and blue circles illustrate the
path $L_1$, purple and orange squares illustrate the path
$L_2$, which are used to define the string operators $\Phi_1$ and $\Phi_2$ in a system with periodic boundary conditions. %(see Eq.~\ref{stringOps}).
%$L_x$ contains the sites colored along a path in the horizontal direction (red and blue circles),
%and $L_y$ contains the sites colored along a path in the vertical direction (purple and orange squares).
}
\end{figure}

\it $Z_n$ Kitaev model--\rm We consider the following Hamiltonian on the honeycomb
lattice with $n$ states per site:
\begin{align}
\label{Znmodel}
H =- \sum_{\langle i j \rangle } J_{s_{ij}} (T_i^{s_{ij}} T_j^{s_{ij}} + H.c.),
\end{align}
where $s_{ij} = x,y,z$ depends on the direction of the link $ij$ (Fig. \ref{fig1}).
$T_{i}^x$ and $T_{i}^y$ {are $n\times n$ matrices} satisfying the relations: $ T_i^x T_i^y = T_i^y T_i^x \omega$, $(T_{i}^x)^n = (T_{i}^y)^n = 1$,
where $\omega \equiv e^{i 2\pi/n}$.% \blue{For example, an explicit choice of matrices $T_i^x,T_i^y$ are $\left[T_i^x\right]_{\alpha\beta}=\delta_{\alpha\beta}e^{i\alpha 2\pi/n},~\left[T_i^y\right]_{\alpha,\beta}=\delta_{\alpha,\beta+1}$. }
We further define: $T_i^z \equiv (T _i^x T_i^y)^\dagger$, which implies
$T_i^z T_i^x = T_i^x T_i^z  \omega$, $T_i^y T_i^z = T_i^z T_i^y \omega$.
$T^s_i$ from different sites commute with each other.
The case $n=2$ corresponds to the original Kitaev model.

The key fact about this model is that there is a conserved operator
associated with each plaquette. Define:
\begin{align}
W_p \equiv \prod_{\langle i j \rangle \in \varhexagon} K_{ij} = (\omega T_1^x T_2^y T_3^z T_4^x T_5^y T_6^z)^\dagger
\end{align}
where the site labels are shown in Fig. \ref{fig1}. %run around a plaquette starting at the bottom left site (see Fig. \ref{fig1}).
Following Kitaev, we define $K_{jk} = T_j^{s_{jk}} T_k^{s_{jk}}$.
%\begin{align}
%K_{jk} = \left\{ \begin{array}{cc}
%T_j^x T_k^x, &\text{ if } (j,k) \text{ is an $x$-link}, \\
%T_j^y T_k^y, &\text{ if } (j,k) \text{ is an $y$-link}, \\
%T_j^z T_k^z, &\text{ if } (j,k) \text{ is an $z$-link}
%\end{array} \right.
%\end{align}
{It can be verified directly that}
%\begin{align}
$[W_p, H] = 0$,
%\end{align}
so that the spectrum can be decomposed into eigenstates of $W_p$. Note that
$W_p^n = 1$.%, which implies that the eigenvalues of $W_p$ are $n$th roots of unity.
%This is the $Z_3$ flux through the plaquettes of the honeycomb model.

In addition to the above conserved plaquette operators, the model (for $n \geq 3$) with
periodic boundary conditions also admits conserved, non-commuting, loop operators:
\begin{align}
\label{stringOps}
\Phi_1 \equiv \prod_{2i-1,2i \in L_1} T^z_{2i-1} T^{z\dagger}_{2i},
\;
\Phi_2 \equiv \prod_{2i-1,2i \in L_2} T^y_{2i-1} T^{y\dagger}_{2i}
%\Phi_y \equiv \prod_{i \in L_y} T_{4i-3}^x T_{4i-2 }^{x \dagger} T^{y \dagger}_{4i-1} T^y_{4i} ,
%\Phi_x &\equiv \prod_{i \in L_x} T^z_{2i-1} (T^z_{2i})^\dagger
%\nonumber \\
%\Phi_y &\equiv \prod_{i \in L_y} T_{4i-3}^x (T_{4i-2}^x)^\dagger (T^y_{4i-1})^\dagger T^y_{4i} ,
\end{align}
where $[\Phi_1,H] = [\Phi_2,H]= 0$, and $\Phi_2 \Phi_1 = \Phi_1 \Phi_2 \omega^{2}$.
The loops $L_1$ and $L_2$ are shown in Fig. \ref{fig1}, and describe non-contractible paths
around the hexagonal lattice in the two directions. Since these operators are conserved,
eigenstates must form a representation of their algebra. This rigorously implies a ground
state degeneracy on the torus that is a multiple of $n$ for $n$ odd, and $n/2$ for $n$ even.

{Just as in the original Kitaev model, the generalized model can be defined on any {\it planar} trivalent graph.
A key difference between the $n\geq 3$ and the $n=2$ cases is that for $n \geq 3$, the three operators
$T_i^{x,y,z}$ on each site must be ordered with the same chirality. In other words, the direction
$x\rightarrow y\rightarrow z\rightarrow x$ must be either all counter-clockwise or all clockwise
on all sites. This requirement also means that the model can only be defined on planar graphs.
Physically, this is because the large loops $\Phi_1,\Phi_2$ defined above can be considered as
Wilson loops of a particle with statistical angle $\frac{2\pi}{n}$. For $n>2$ this particle is an
Abelian anyon, which can only be defined in two-dimensions, while for $n=2$ it is a fermion.}
Multi-site terms can be added {to the Hamiltonian} without affecting the conservation laws,
{as long as they are products of bond terms $K_{ij}$ and/or $K_{ij}^\dagger$.
%In the Supplemental Materials,
In the supplementary materials\cite{supplement}, we present more details of the computation
of commutation relations and conserved quantities by setting up convenient diagrammatic rules. }

\it Anisotropic limit and the Abelian phase \rm-- Similar to the original model\cite{kitaev2006},
the anisotropic limit $J_z \gg J_x, J_y$ can be easily solved. In this limit, we first diagonalize the $J_z$ terms in the Hamiltonian.
To do this, let us pick a basis of $n$ states on each site, $|a\rangle_i$, which diagonalize
$T^z_i$: $T^z_i|a \rangle_i = \omega^a |a \rangle_i$, for $a = 0,...,n-1$.
Pairs of sites $i$, $j$ coupled by $J_z$ have their $n^2$ states split into
$n$ degenerate lowest energy states, $|a\rangle_i |n-a\rangle_j$, for
$a = 0, .., n-1$. These states are separated by a gap of order $J_z$ relative
to the remaining $n^2-n$ states. For large $J_z$, we can treat
pairs of sites separated by vertical links effectively as a single
site, thus obtaining at low energies a square lattice with
$n$ states per site. Within the degenerate $n$-dimensional space on
each site, we can define a new set of $Z_n$ rotor operators $L^x_{i}$, $L^y_{i}$, such that
$L^x_{i} |a\rangle_i |n-a\rangle_j = \omega^a |a \rangle_i |n-a\rangle_j$,
and $L^y_{i} |a\rangle_i |n-a\rangle_j = |a-1\rangle_i |n-a+1\rangle_j$.

Within this low-energy subspace, the remaining $J_x$ and $J_y$ terms can be treated
within perturbation theory. The lowest order term that does not change the $J_z$ bond energy is
$\frac{J_x^2 J_y^2}{(6 J_z)^3} K_{12} K_{23} K_{45}K_{56}$ {(with the label of sites defined in Fig. \ref{fig1}).}
It is straightforward to show that this gives
$H_{eff} = \frac{J_x^2 J_y^2}{(6 J_z)^3}  \sum_{ijkl \in \square} L_i^x L_j^y L_k^x L_l^y$, which is
the $Z_n$ toric code Hamiltonian \cite{kitaev2003,wen2003,schulz2012}.
\begin{figure}
\centerline{
\includegraphics[width=3in]{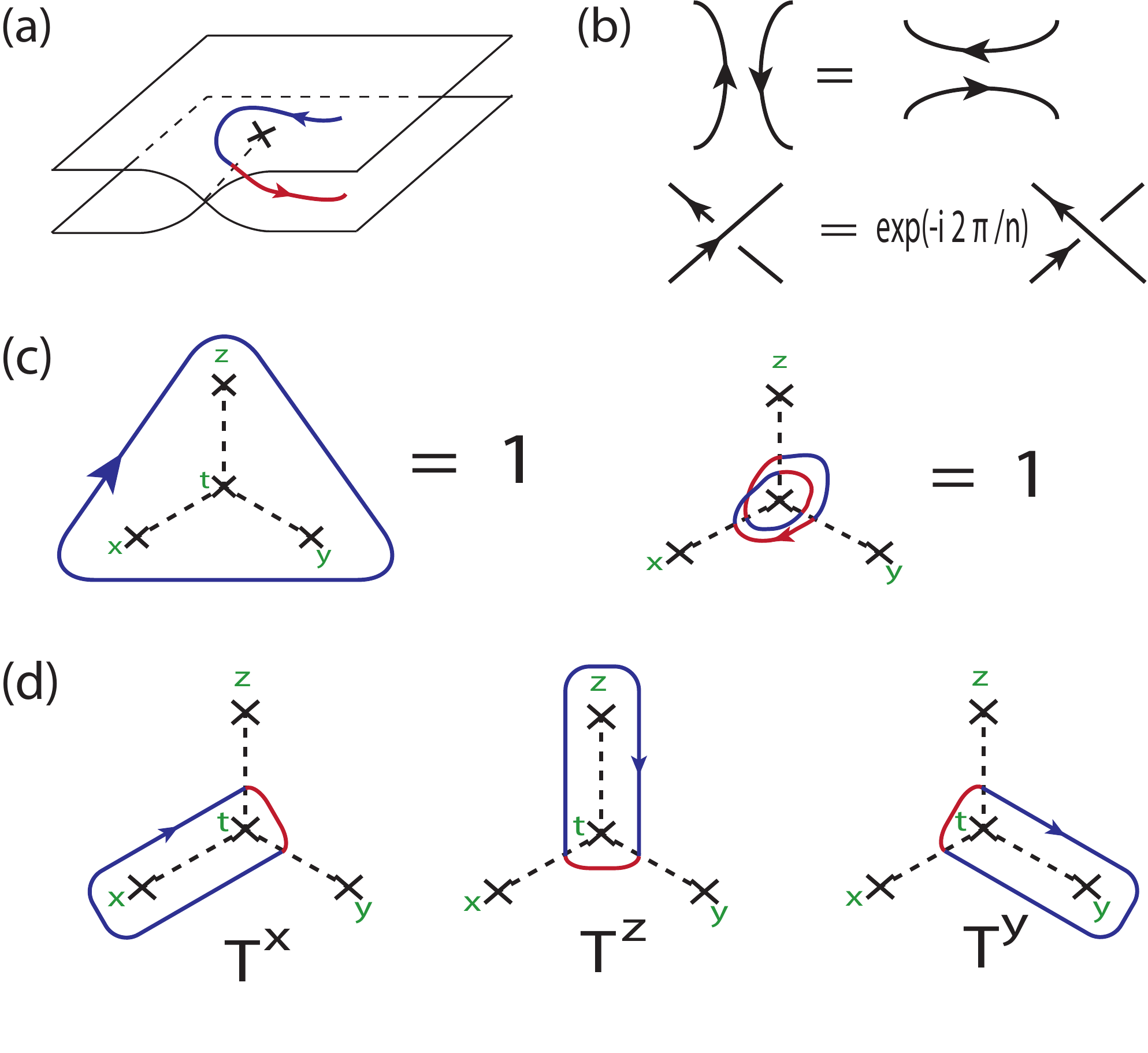}
}
\caption{
\label{slaveGenon}
(a) A genon (twist defect) in a bilayer FQH system, is marked by the X. The branch cut emanating
from the genon connects the two layers. (b) The Wilson lines of the
Abelian quasiparticles
can locally be cut and rejoined.
 (c) A spin is represented in terms of $4$ genons, labelled $x$, $y$, $z$, $t$.
The constraint that the Wilson loop around all 4 genons be trivial reduces the number of states to
$n$. The double loop around each genon can always  be set to one.
(d) $T^x$, $T^y$, $T^z$ correspond to Wilson loop operators
around pairs of genons.
}
\end{figure}

\it Slave Genons \rm-- In order to further analyze the model beyond this strongly anistropic
limit, we introduce a `slave genon' approach, {which maps the spin model to a model of
coupled non-Abelian twist defects \cite{barkeshli2012a,barkeshli2013genon,barkeshli2010,bombin2010,
kitaev2012,you2012,lindner2012,clarke2013,cheng2012,vaezi2013,teo2013,brown2013,barkeshli2013,khan2014}, referred to as genons
\cite{barkeshli2012a,barkeshli2013genon}, in a topologically ordered state.} This generalizes the Majorana fermion representation
introduced in the original Kitaev honeycomb model \cite{kitaev2006}, along with well-known slave
fermion/boson techniques \cite{wen04}. {A key difference in the $n\geq 3$ $Z_n$ models is that the slave particles must
be non-local topological defects instead of fermions or bosons.}

{Consider a Laughlin $1/n$ fractional quantum Hall (FQH) state on the surface shown in Fig. \ref{slaveGenon} (a). The surface is obtained by introducing a branch cut line in a bilayer system, such that the two layers are exchanged across the branch cut line. A genon is defined as the endpoint of the branch cut line\cite{barkeshli2010,barkeshli2012a,barkeshli2013genon}.
Now consider $4$ genons with the constraint that they fuse to vacuum. As is shown in Fig. \ref{slaveGenon} (c), this constraint means a Laughlin quasparticle going around the $4$ genon cluster obtains no Berry's phase. With this constraint, the disk region with $4$ genons is topologically equivalent to a torus with a single layer of $1/n$ state\cite{barkeshli2010}, which thus has $n$ topological ground states. The slave genon approach is defined by mapping the $n$-state rotor on each site of the honeycomb lattice to such a cluster of $4$ genons.}
{The spin operators $T_i^{x,y,z}$ are mapped to Wilson loop operators, defined as the unitary rotation of topological ground states induced by adiabatic propagation of charge $1/n$ Laughlin quasiparticles along a non-contractible loop. $T_i^{x,y,z}$ corresponds to the three non-contractible loops shown in Fig. \ref{slaveGenon} (d).
During topological deformations of the Wilson loops, we also require that a double loop around a genon is contractible,
as is illustrated in \ref{slaveGenon} (c). Physically this removes the ambiguity that a genon may trap a Laughlin quasiparticle.}
%  The operators for different non-contractible loops do not commute with each other, which yield topological degeneracies.
We emphasize that the genons and associated FQH state are \it entirely auxiliary \rm degrees of freedom --
the spin model is not required to have a FQH state physically.

\begin{figure}
\centerline{
\includegraphics[width=3in]{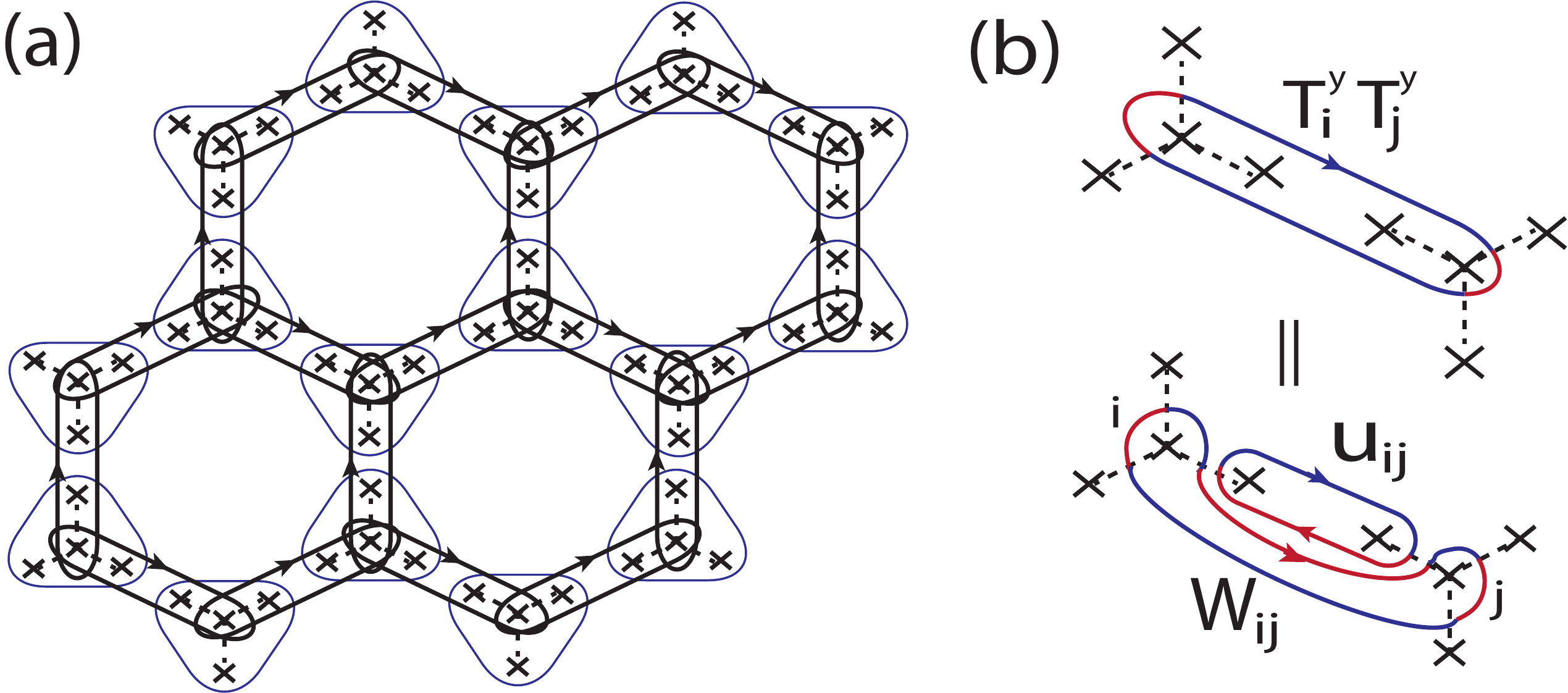}
}
\caption{\label{gauge}
(a) The interaction terms in the Hamiltonian correspond to the three types of loops. The blue loop around each site represents the local constraint which commute with the Hamiltonian terms. 
(b) A loop corresponding to the interaction $T_i^y T_j^y$ can be decomposed into two
non-overlapping loops, $W_{ij}$ and $u_{ij}$.
}
\end{figure}

{In this representation, the spin model is mapped to a two-dimensional array of genons, with couplings
given by Wilson loop operators. The two-site terms $K_{ij}$ in the Hamiltonian simply correspond
to Wilson loops surrounding $4$ genons, as is shown in Fig. \ref{gauge}. Importantly, the
Hamiltonian commutes with the local constraint at each site, since the Wilson loop corresponding
to the local constraint commutes with that of $K_{ij}$, as is illustrated in Fig. \ref{gauge} (a).
On each site, the constraint can be expressed in the spin operators $T_i^{x,y,z}$ as
$D_i\equiv T_i^xT_i^yT_i^z=1$, which projects the $n^2$ states of $4$
genons\cite{barkeshli2010} to $n$ states of the physical spin. }

From the pictorial representation, we readily infer that the Hamiltonian can be rewritten as:
\begin{align}
\label{parafArray}
H = -\sum_{\langle ij \rangle} J_{s_{ij}} u_{ij} W_{ij} + H.c. ,
\end{align}
where $W_{ij}$ and $u_{ij}$ are the loop operators corresponding to the operation
of moving charge $1/n$ Laughlin quasiparticles around the loops shown in Fig. \ref{gauge}b.
Note that $u_{ij}$ only appears in the Hamiltonian in the term $T_i^{s_{ij}} T_j^{s_{ij}}$. From Fig. \ref{gauge}, we deduce that $[u_{ij}, W_{ij}] = 0$, {$\left[u_{ij},u_{kl}\right]=0$} and therefore $[u_{ij}, H] = 0$. We can hence
replace the $u_{ij}$ by $c$-numbers, associated with different superselection sectors. {$W_{ij}$
can be considered as a two-dimensional ``parafermion hoping" term, while the eigenvalues
of $u_{jk}$ can be considered as a $Z_3$ gauge field coupled to the parafermions \cite{fendley2012}.
The precise meaning of the parafermion coupling will be discussed in next paragraph. } 
By deforming the loops $u_{ij}$ and using the constraints shown in Fig. \ref{slaveGenon}c,
it is straightfoward to show that the conserved plaquette operators, $W_p$, correspond to
the $Z_n$ ``gauge flux'' through a plaquette:
%\begin{align}
$W_p = \prod_{\langle ij \rangle \in \hexagon } u_{ij} = u_{12} u_{23} u_{34} u_{45} u_{56} u_{61}$.
%\end{align}

\begin{figure}
\centerline{
\includegraphics[width=2.3in]{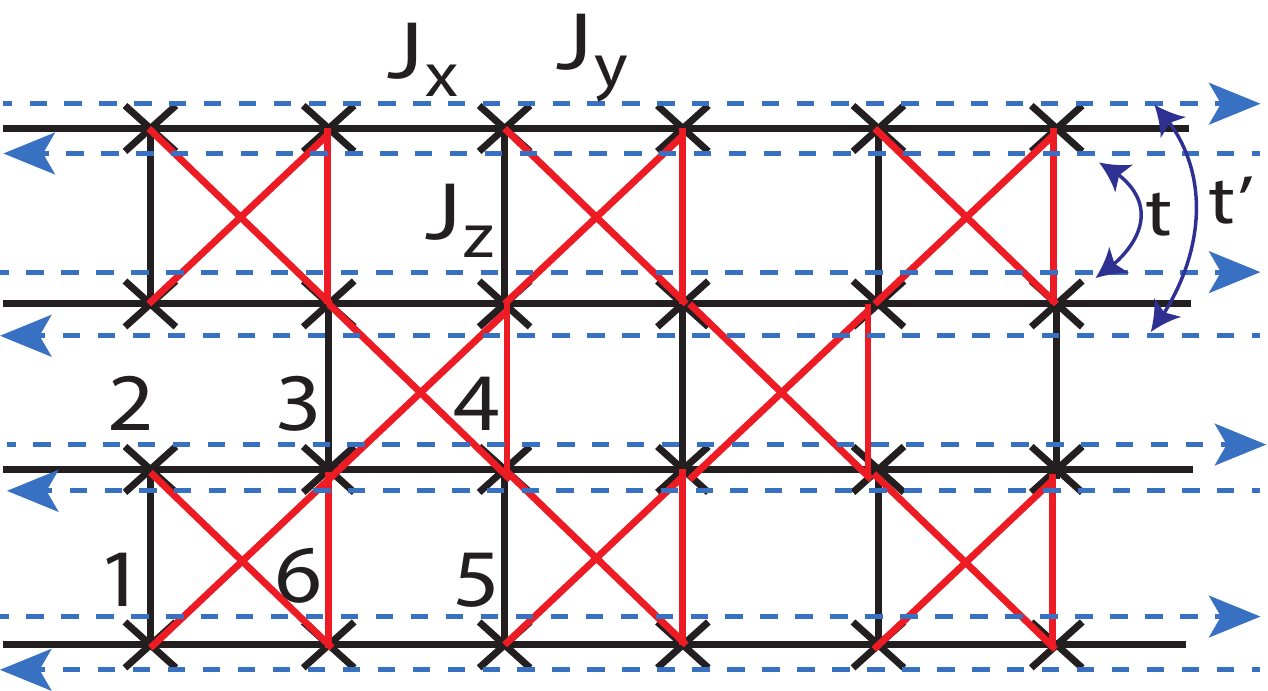}
}
\caption{Hamiltonian (\ref{parafArray}) describes a hexagonal array of
  coulpled genons, or $Z_n$ parafermion
zero modes. For $J_x = J_y$, and in the absence of interchain
interactions, each chain is at criticality, which in the $n=3$ case is described by a
$Z_3$ parafermion CFT. Interchain coupling terms can be added to gap out counterpropagating
parafermion modes from each chain, leading to a gapped topologically ordered state with a chiral
$Z_3$ parafermion edge mode. Red bonds correspond to the next neighbor interactions (see (\ref{modifiedH})).
 \label{coupledchain}}
\end{figure}

{To understand more explicitly the meaning of coupled parafermion zero modes, we} %To see this more explicitly, let us
first consider the Hamiltonian for a single chain, with $u_{ij}$ uniformly set to 1:
$H_{1D} = -\sum_{i} (J_x W_{2i-1,2i} + J_y W_{2i,2i+1} + H.c.)$, with $W_{i-1,i} W_{i,i+1} = W_{i,i+1} W_{i-1,i} \omega$.
This Hamiltonian is equivalent to the transverse field $Z_n$ Potts
model. Following the results in the Potts model \cite{fradkin1980,fendley2012,fendley2014}, a pair of parafermion
operators $\alpha_{Li},~\alpha_{Ri}$ can be introduced, which satisfies the algebra
$\alpha_{R/Li} \alpha_{R/Lj} = \alpha_{R/Lj} \alpha_{R/Li} e^{\pm i 2\pi sgn(j-i)/n}$.
In terms of spin operators of the Kitaev model, we have
$\alpha_{Ri}=T_{1y}^\dagger K_{12}K_{23}...K_{i-1,i}$, $\alpha_{Li}=
T_{1y}^\dagger K_{12}^\dagger K_{23}K_{34}^\dagger...K_{i-1,i}^{s_i}$,
with $s_i=-(-1)^i$. $H_{1D}$ can be rewritten in terms of a ``parafermion chain'' by setting
$W_{i,i+1} \propto \alpha_{Ri}^\dagger \alpha_{R,i+1}^{\phantom{\dagger}}$.
The 2D Hamiltonian $(\ref{parafArray})$ can then be reinterpreted as an array of coupled
1D parafermion chains~\cite{fendley2012,burrello2013,bondesan2013,mong2013,vaezi2013b,supplement}.

{The single chain system with $n=3$ is particularly interesting.} When $J_x = J_y$, the model is at a self-dual critical point of the 1D $Z_3$ Potts model,
which is known to be described by a $Z_3$ parafermion conformal field theory (CFT) {with central charge $c=4/5$}\cite{zamolodchikov1985}. {At small but finite $J_z$, the system can be viewed as coupled parafermion chains, as is illustrated in Fig. \ref{coupledchain}. It is known that a ``chiral" coupling between 1D gapless chains can realize a chiral 2D topologically ordered state\cite{yakovenko1991,mong2013,
vaezi2013b,titus1403}, if the right-moving (left-moving) states of a chain are only coupled
to the left-moving (right-moving) states of the chain below (above) by a relevant coupling. In our $n=3$ system, such a
coupling, if realized, will result in a non-Abelian topological state with chiral $Z_3$ parafermion
edge states. This is similar to the proposal of \cite{mong2013} although the latter is not a local
spin model and therefore realizes a different topological order. In $n>2$ models, the $J_z$ coupling
breaks time-reversal symmetry, so that it is possible for the system with some proper $J_z$ to
be in the same non-Abelian phase as the ideal system with only chiral coupling.}

\begin{figure}
\centerline{
\includegraphics[width=2.7in]{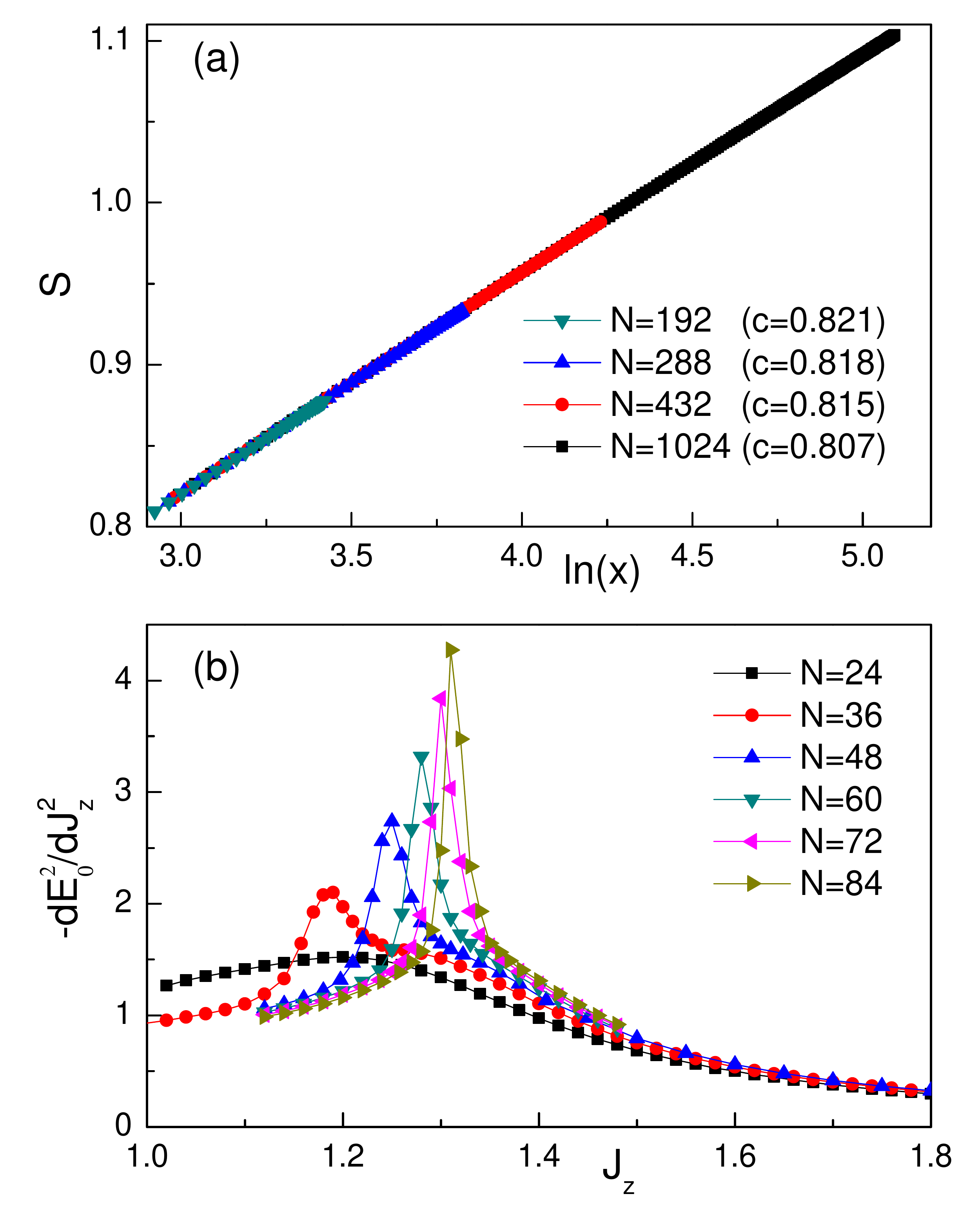}
}
\caption{\label{numerics}
(a) Entanglement entropy of an open $N$ site chain. The fit to $S(x) = \frac{c}{6}\ln(x)+const$ extrapolates to a central charge $c = 4/5$, where $x=\frac{N}{\pi}\sin(\frac{\pi l}{N})$ and $l$ is subsystem length. (b) Second derivative of the ground state energy density as a function of the distortion $J_z$, computed
from DMRG with 3 chains. A phase transition is resolved when $J_z$ is tuned between the isotropic
and anisotropic limits.}
\end{figure}

\it Numerical Results \rm-- { To gain further understanding of the $n=3$ system}, we have performed preliminary numerical analyses. 
For the single chain with $J_x=J_y$, our DMRG results~\cite{White1992DMRG,JiangBalents} for the entanglement entropy shows that the chain is indeed described by a conformal field theory
with central charge $c = 4/5$, as shown in Fig. \ref{numerics} (a). 
In the opposite limit $J_z \gg J_x, J_y$, we have verified through exact diagonalization that the system is gapped, with a $9$-fold ground state degeneracy. As $J_z$ is lowered relative to $J_x, J_y$, we expect a phase transition from the Abelian phase to the isotropic phase. Fig. \ref{numerics}(b) shows DMRG results for the second derivative of the ground state energy density, $-d^2 E_0/d J^2_z$, which indeed shows evidence of a sharp phase transition. These numerical results confirm non-trivial features of the
$Z_3$ Kitaev model, while they do not fully establish the nature of the isotropic phase. {More complete numerical study of the non-Abelian phase will be left for future works.}%We will present a more detailed numerical analysis elsewhere.

\it {Multi-site terms and the} controlled limit\rm-- %The transformed problem, eq. (\ref{parafArray}), is still a non-trivial, interacting problem, which can be thought of as an array of interacting parafermion zero modes. Nevertheless, specializing to the case $n=3$, we will consider a modification of this Hamiltonian that is analytically tractable. As is illustrated in Fig. \ref{coupledchain}, in the $J_x=J_y, J_z\rightarrow 0$ limit the system can be viewed as weakly coupled $Z_3$ parafermion CFT's.
{In the original Kitaev model\cite{kitaev2006}, a three site term drives the model into the non-Abelian Ising phase. %In the coupled chain analysis above, the role of this term is to introduce chiral coupling between neighboring chains.
Similarly, for $n=3$ it is possible to consider a modification of the Hamiltonian (\ref{Znmodel}) that
makes the non-Abelian state more tractable. As is pointed out in Ref. \cite{mong2013,vaezi2013c}, there is a
known correspondence between the lattice parafermion operators and continuous fields in the
$Z_3$ Potts model CFT. Using this correspondence, one can see that the parafermion coupling
of the form $-\lambda \sum_{j,m} \left(\alpha_{R,2j,m}^\dagger +\alpha_{R,2j+1,m}^\dagger\right)\left(\alpha_{L,2j,m+1}+\alpha_{L,2j+1,m+1}\right)+h.c.$ between two neighboring chains labelled by $m$ and $m+1$ induces the chiral coupling between the right movers of the $m$-th chain and the left movers of the $m+1$-th chain. Since this is a direct application of Ref. \cite{mong2013,vaezi2013c}'s result, we will
leave more detailed derivation of this term for the supplementary materials\cite{supplement}.}

{Using the Wilson loop representation, the chiral coupling between parafermions reviewed above can be achieved in
a local spin Hamiltonian: }
\begin{align}
\label{modifiedH}
H' = H - J_z \sum_{\hexagon} \mathcal{O}_{\hexagon},
\end{align}
with $\mathcal{O}_{\hexagon} = (T^z_1 T^y_1 T^z_2  T^y_6
+ T^x_3 T^x_2 T^z_2 T^z_1 T^y_1 T^y_6 + T^x_3 T^x_2 T^z_2 T^z_1  + H.c.)$, and $H$ given by Eq. (\ref{Znmodel}). Therefore,
the above Hamiltonian, with $J_x = J_y \gg J_z > 0$, could realize a gapped, 2D topologically ordered state,
with a robust chiral $Z_3$ parafermion CFT propagating along its boundary. The topological order can then
be read off from the primary field content of the $Z_3$ parafermion CFT \cite{zamolodchikov1985,difrancesco}.

We emphasize that the possibility of realizing a coupled array of parafermion chains, with couplings
that involve only single parafermion operators from different chains, is highly non-trivial. This is
not possible with the usual transverse field Potts model, but is possible with the
approach described here. The slave genon transformation thus provides a way to
design general interactions in 2D lattices of parafermions, in terms of \it local \rm interactions of a
2D spin model. We expect a similar method can be employed for {much more general models, which may enable a spin model realization of the anyon lattice models studied in the literature
\cite{adrian2007,gils2009}}.

\it Further generalizations \rm--{ The model described here admits much more generalization. For example, one can consider genons in a generic Abelian FQH state. Quasiparticles in each layer are labeled by integer vectors $\vec{l}$, with the
fractional mutual statistics $\theta_{ll'} = 2\pi \vec{l}^T K^{-1} \vec{l'}$ and self statistics
$\theta_l = \pi \vec{l}^T K^{-1} \vec{l}$ determined by an integer valued $K$ matrix\cite{wen04}. $4$
genons with the local constraint in Fig. \ref{slaveGenon} now correspond to a spin with
$|K|$ states\cite{barkeshli2013genon}. The spin operators $T_i^{x,y,z}$ generalize to Wilson loop operators $T^{x,y,z}_{\vec{l}}$ of a quasiparticle $\vec{l}$ around the same loops as those in Fig. \ref{slaveGenon} (d). }
These operators satisfy the algebra
%\begin{align}
$T^x_{\vec{l}} T^y_{\vec{l}'} = T^y_{\vec{l'}} T^x_{\vec{l}} e^{i 2\pi \vec{l}^T K^{-1} \vec{l}'}$,
%\end{align}
$T^s_{\vec{l}} T^s_{\vec{l}'} = T^s_{\vec{l} + \vec{l}'}$, and $T_{K\vec{n}} = 1$ for all $\vec{n}\in\mathbb{Z}^N$.
Therefore, we can consider the more general Kitaev-type Hamiltonian:
$H = \sum_{\vec{l} \in \mathbb{Z}^N} \sum_{\langle i j \rangle} J_{\vec{l}; s_{ij}} T_{\vec{l}}^{s_{ij}} T_{\vec{l}}^{s_{ij}} + H.c.$
This model can be analyzed similarly to the $Z_n$ generalization presented earlier. In particular,
there are conserved quantities associated with each plaquette {and conserved large loop operators on torus geometry}, and one can consider an exact
transformation to a lattice model of interacting genons or, alternatively, generalized parafermion
zero modes\cite{barkeshli2013defect}.

\noindent{\bf Acknowledgement.} {The $Z_n$ generalization of the Kitaev model has also been
studied independently in unpublished works of P. Fendley and C. L. Kane ({\it c.f.} discussions in \cite{fendley2012}).
This work was presented in June of 2013 on a workshop\cite{simonstalk}.
As this manuscript was being completed, we were made aware that a related but different generalization of Kitaev model has been studied by A. Vaezi\cite{Vaezi}.
}
We thank A. Ludwig and A. Kitaev for discussions.
HCJ was supported by the Templeton Fund. RT has been supported 
by the European Research Council through
ERC-StG-TOPOLECTRICS-336012. XLQ is 
supported by David \& Lucile Packard foundation. We also acknowledge
computing support from the Center for Scientific Computing
at the CNSI and MRL: NSF MRSEC (DMR-1121053) and
NSF CNS-0960316.

%\bibliographystyle{prsty}
%\bibliography{TI}

\newpage
\newpage
\begin{widetext}

\section{Supplementary material for ``Generalized Kitaev Models and Slave Genons"}
%\author{Maissam Barkeshli}
%\affiliation{Microsoft Station Q, Santa Barbara, CA, 93106, USA}
%\author{Hong-Chen Jiang}
%\affiliation{Department of Physics, University of California, Berkeley, California 94720, USA}
%\author{Ronny Thomale}
%\affiliation{Institute for Theoretical Physics, University of W\"urzburg, Am Hubland, D-97074 W\"urzburg, Germany}
%\author{Xiao-Liang Qi}
%\affiliation{Department of Physics, Stanford University, Stanford, CA 94305 }

%\maketitle

\section{Construction of the $Z_n$ Kitaev model on more general lattices}

In this section, we will explain the general rules we use to define the $Z_n$ Kitaev model and obtain the conserved quantities. These rules can then be generalized to define these models on generic trivalent lattices. We start with the $Z_n$ algebra
\begin{eqnarray}
T_i^xT_i^y=T_i^yT_i^x\omega,~T_i^yT_i^z=T_i^zT_i^y\omega,~T_i^zT_i^x=T_i^xT_i^z\omega
\end{eqnarray}
with $\omega=e^{i 2\pi /n}$. This algebra can be summarized by drawing
a triangle on each site of the honeycomb lattice, as is shown in
Fig. \ref{fig:supp1} (a). The three vertices of the triangle represent
the three operators $T_i^{x,y,z}$, with the arrow an indicator of
their commutation relation. For any two of the three operators
$T_i^\alpha,~T_i^\beta$, $T_i^\alpha T_i^\beta=T_i^\beta T_i^\alpha
\omega$ if $\alpha\rightarrow \beta$ is along the arrow direction,
while the phase factor is $\omega^{-1}$ if $\alpha\rightarrow \beta$
is the reverse of the arrow direction. This arrow rule will be helpful when we check the commutation relation between terms in the Hamiltonian and define the conserved quantities.

Consider the Hamiltonian given by Eq. (1) of the manuscript. We denote the term on each bond as $K_{ij}=T_i^{s_{ij}}T_j^{s_{ij}}$, with $s_{ij}=x,y,z$ depending on the bonds. Consider the plaquette formed by sites 1,2,...,6 shown in Fig. \ref{fig:supp1} (b). Using the arrow rule, and remembering that the spins on different sites commute with each other, we can easily check that
\begin{eqnarray}
K_{12}K_{61}=K_{61}K_{12}\omega,~K_{12}K_{23}=K_{23}K_{12}\omega^{-1}
\end{eqnarray}
This is simply  because the arrow goes out of bond $12$ at site $1$, while it goes into the bond at site $2$. In the plaquette operator
 \begin{eqnarray}
 W_p=K_{12}K_{23}K_{34}K_{45}K_{56}K_{61}
 \end{eqnarray}
 $K_{16}$ and $K_{23}$ are the only two terms which do not commute with $K_{12}$. Therefore we see that the factor given by the two terms cancel, and we obtain
 \begin{eqnarray}
 \left[K_{12},W_p\right]=0
 \end{eqnarray}
This proves that the Hamiltonian commutes with $W_p$, and also proves that the plaquette operators $W_p$ commute with each other, since they are products of $K_{ij}$.
\begin{figure}
\centerline{
\includegraphics[width=4in]{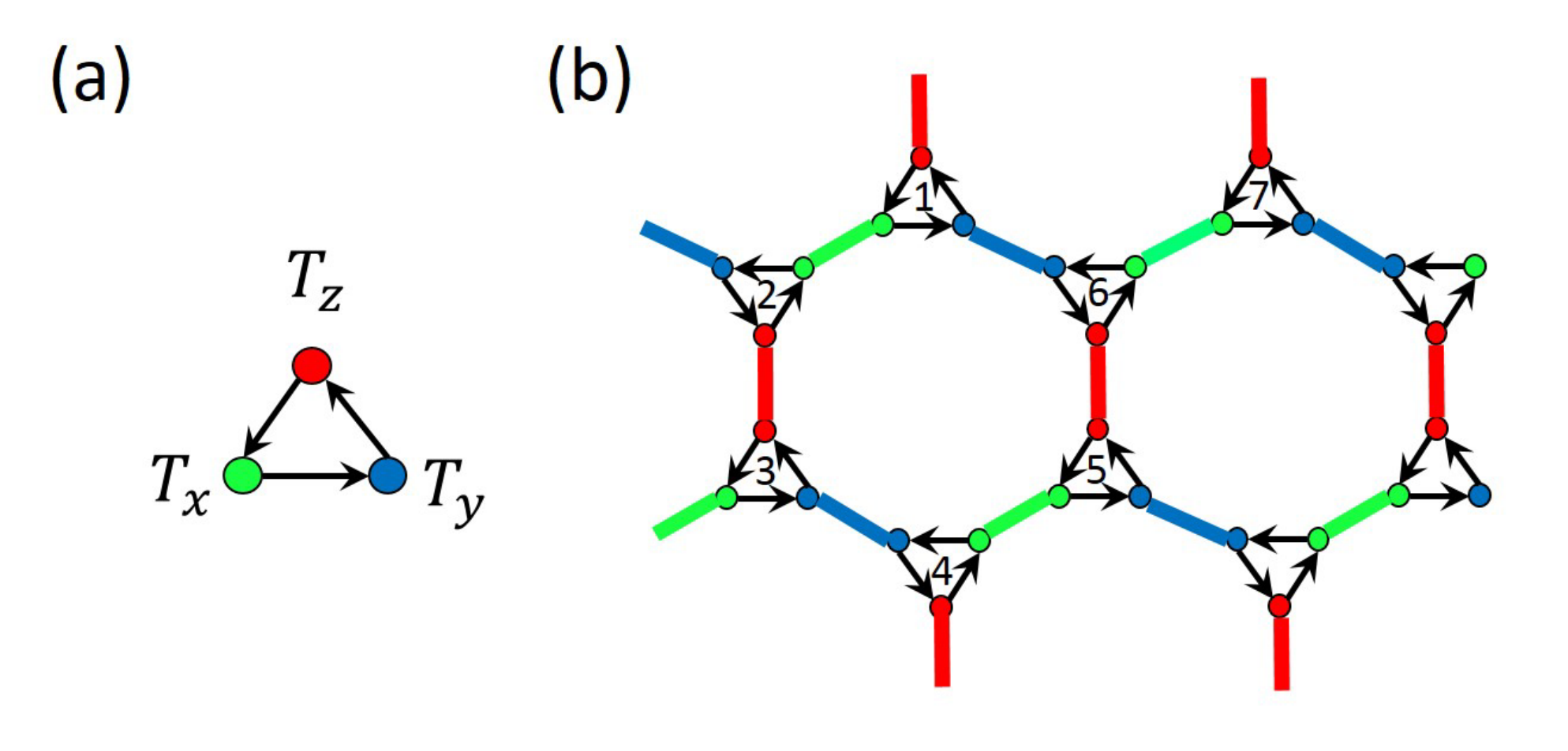}
}
\caption{
\label{fig:supp1}
(a) The arrow between three operators on one site, indicating their commutation relations (see text); (b) The arrows drawn on each site of a honeycomb lattice, which is used to determine the commutation relation between operators in the Hamiltonian.
}
\end{figure}

Compared with the $Z_2$ Kitaev model, in the $Z_n$ model with $n>2$ a new requirement needs to be satisfied in order for the plaquette conserved quantities to be defined: the arrows at each site must have the same chirality. In the choice we make, the arrows go around the triangle in a counter-clockwise order. If the arrow is reversed on some site (which can be done by replacing $T^\alpha_i\rightarrow T_i^{\alpha \dagger}$ only on that site), the two arrows connected to a given bond will be both out or both in, which makes the bond operator $K_{ij}$ non-commuting with $W_p$.

\begin{figure}
\includegraphics[width=5.5in]{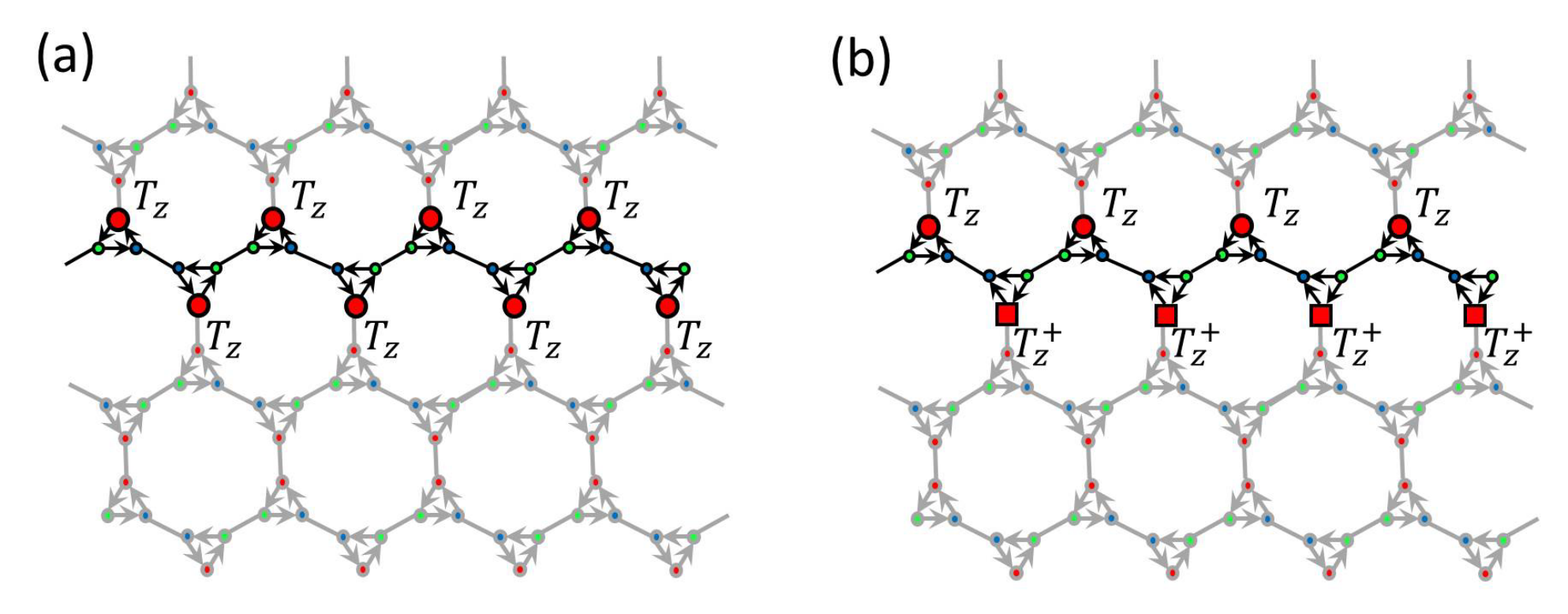}\\
\includegraphics[width=5.5in]{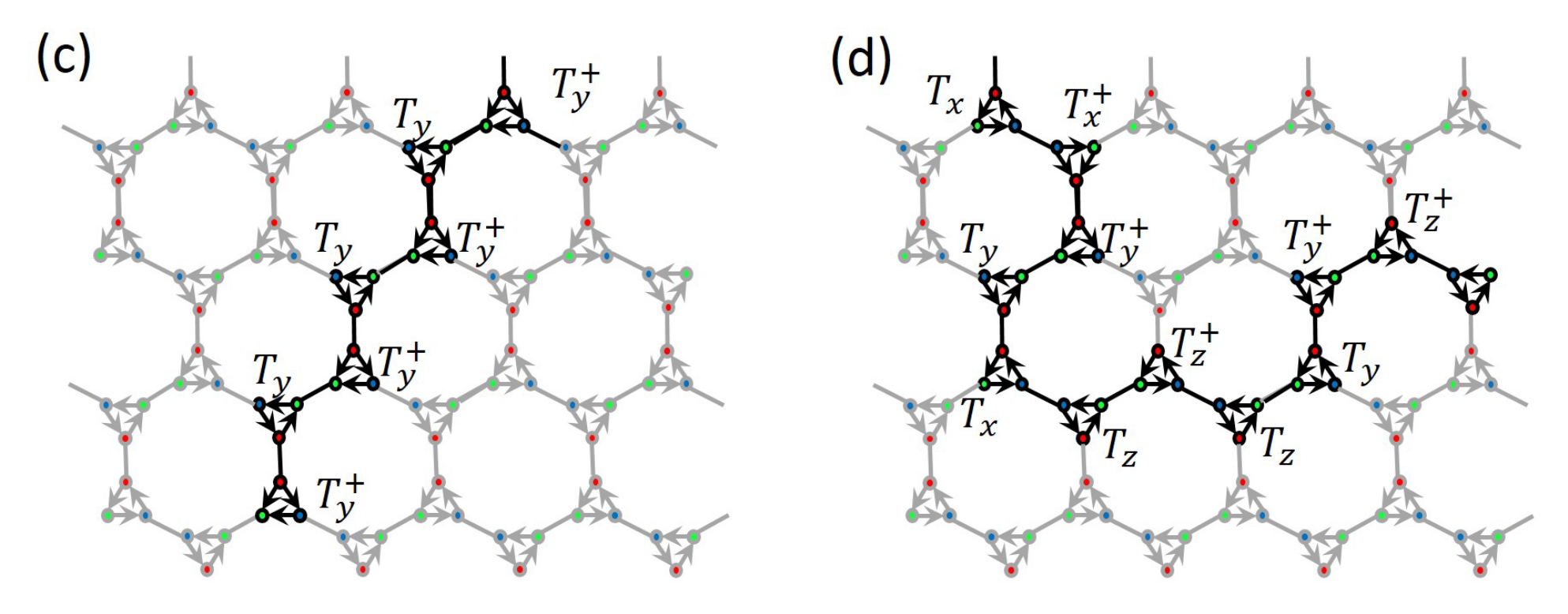}
\caption{
\label{fig:supp2}
(a) The naive construction of large loop operator $\tilde{\Phi}_1$ which does not commute with bond terms $K_{ij}$. (b) The correct large loop operator $\Phi_1$ which commutes with every $K_{ij}$. (c) Another large loop operator along a different direction. (d) A generic loop operator, which is a product of $T_i^\alpha$ at all right-turn corners, and $T_i^{\alpha \dagger}$ at all left-turn corners.
}
\end{figure}

Now we define the large loop operators in a system defined on the torus. In the $Z_2$ Kitaev model, the large loop operator can be obtained by multiplying bond terms $K_{ij}$. In the $Z_n$ case, if we follow this definition and define, for example, 
\begin{eqnarray}
\tilde{\Phi}_1=\prod_{i\in L_1}K_{i,i+1}=\omega^N\prod_{i\in L_1}T_i^{z\dagger}
\end{eqnarray}
as the product of $K_{ij}$ along the zigzag line $L_1$ shown in Fig. \ref{fig:supp2} (a) (where $L_1$ contains $2N$ sites), 
it does not commute with $K_{16}$, since
\begin{eqnarray}
K_{16}K_{12}=K_{12}K_{16}\omega^{-1}, ~K_{16}K_{67}=K_{67}K_{16}\omega^{-1}.
\end{eqnarray}
To define the correct large loop operator, we can change the arrow direction every other site by replacing $T_{2i}^z$ by $T_{2i}^{z\dagger}$ on all even sites. This results in the loop operator shown in Fig. \ref{fig:supp2} (b), defined as
\begin{eqnarray}
\Phi_1=\prod_{2i-1,2i\in L_1}T_{2i-1}^{z}T_{2i}^{z\dagger}
\end{eqnarray}
which can be verified as commuting with the Hamiltonian. Following this rule, another large loop operator can be defined on the other large loop $L_2$ of the torus, as is shown in Fig. \ref{fig:supp2} (c), and written in Eq. (3) of the main text. Interestingly, the two large loops do not commute with each other, as has been discussed in the main text.

More generally, a loop operator can be defined for any given loop $L$ drawn on the honeycomb lattice, as is illustrated in Fig. \ref{fig:supp2} (d). The general rule is the following. i) For each site $i\in L$, pick $\alpha_i$ ($\in x,y,z$) to be the bond type of the bond that is not included in $L$. ii) We define an orientation of the loop. Following this orientation, define a number $s_i=\pm 1$ for each vertex $i$, such that $s_i=1$ or $-1$ if the loop turns right or left at the vertex, respectively. Then the loop operator is defined as
\begin{eqnarray}
\Phi_L=\prod_{i\in L}\left[T_{i}^{\alpha_i}\right]^{s_i}
\end{eqnarray}
(It should be remembered that $\left[T_i^{\alpha}\right]^{-1}=T_{i}^{\alpha\dagger}$. )

Using the arrow rules discussed above, the $Z_n$ Kitaev model can be generalized to all {\it planar} trivalent lattices. At each vertex, the operators $T_i^{x,y,z}$ can be assigned to the three bonds connecting this site, with the order of $x,y,z$ bonds following the same chirality at each site. Denote the spin operator assigned to a bond $ij$ at site $i$ as $T_i^{\alpha_{ij}}$, we can write the Hamiltonian as
\begin{eqnarray}
H=\sum_{\left\langle ij\right\rangle}\left[J_{ij}T_i^{\alpha_{ij}}T_j^{\alpha_{ji}}+h.c.\right]
\end{eqnarray}
In general, $\alpha_{ij}\neq \alpha_{ji}$. An example of a different trivalent lattice and the corresponding operator assignment is shown in Fig. \ref{fig:supp3}. The plaquette conserved quantities and the generic loop operators can all be defined in the same way as on honeycomb lattice.
\begin{figure}
\includegraphics[width=5in]{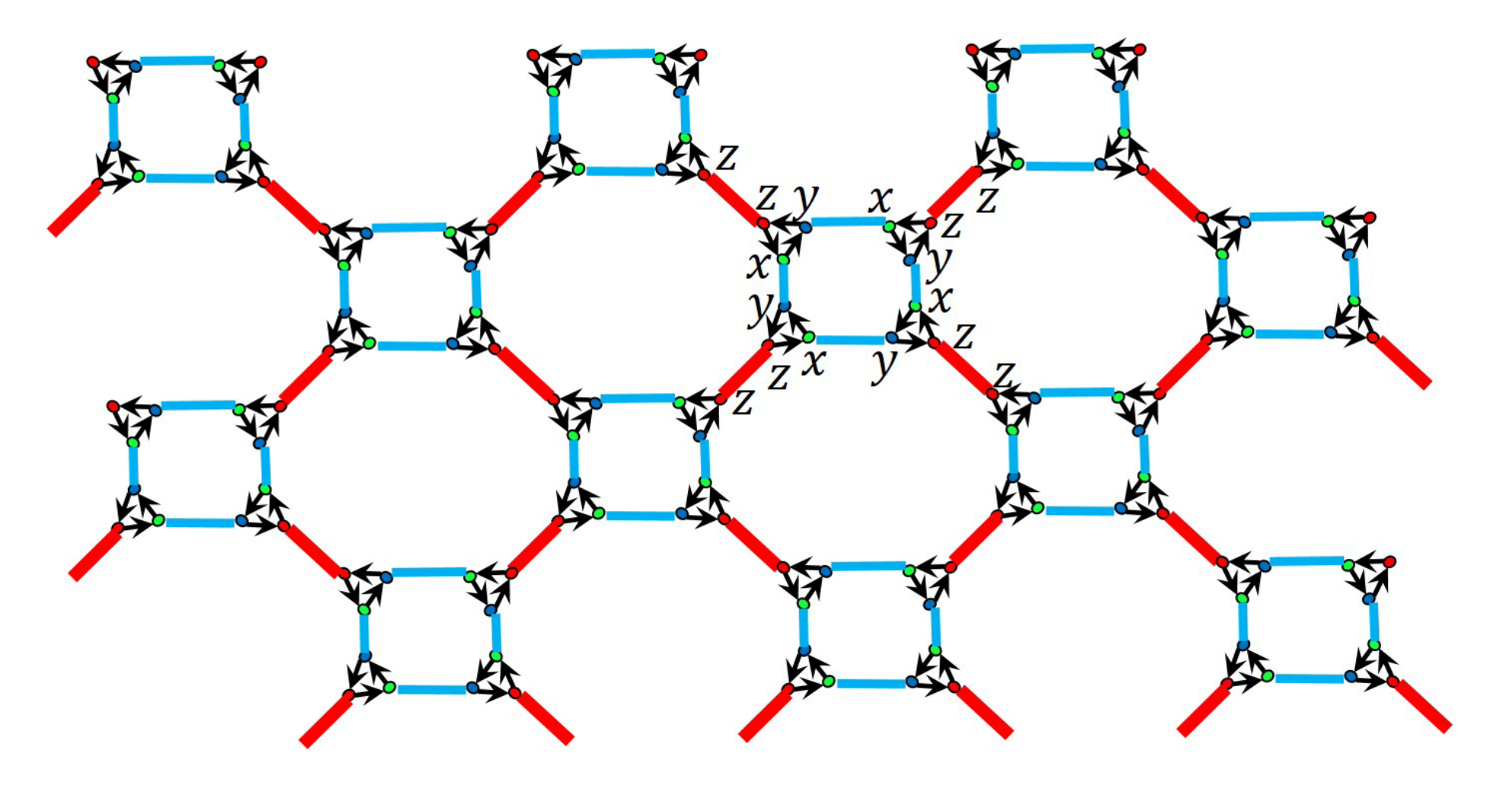}
\caption{
\label{fig:supp3}
Definition of the $Z_n$ Kitaev model on the square-octahedron lattice. The terms in the Hamiltonian are $J_zT_{i}^zT_{j}^z+h.c.$ on the red bonds, and $J_{xy}T_{i}^xT_{j}^y+h.c.$ on the blue bonds. 
}
\end{figure}

\begin{figure}
\includegraphics[width=5in]{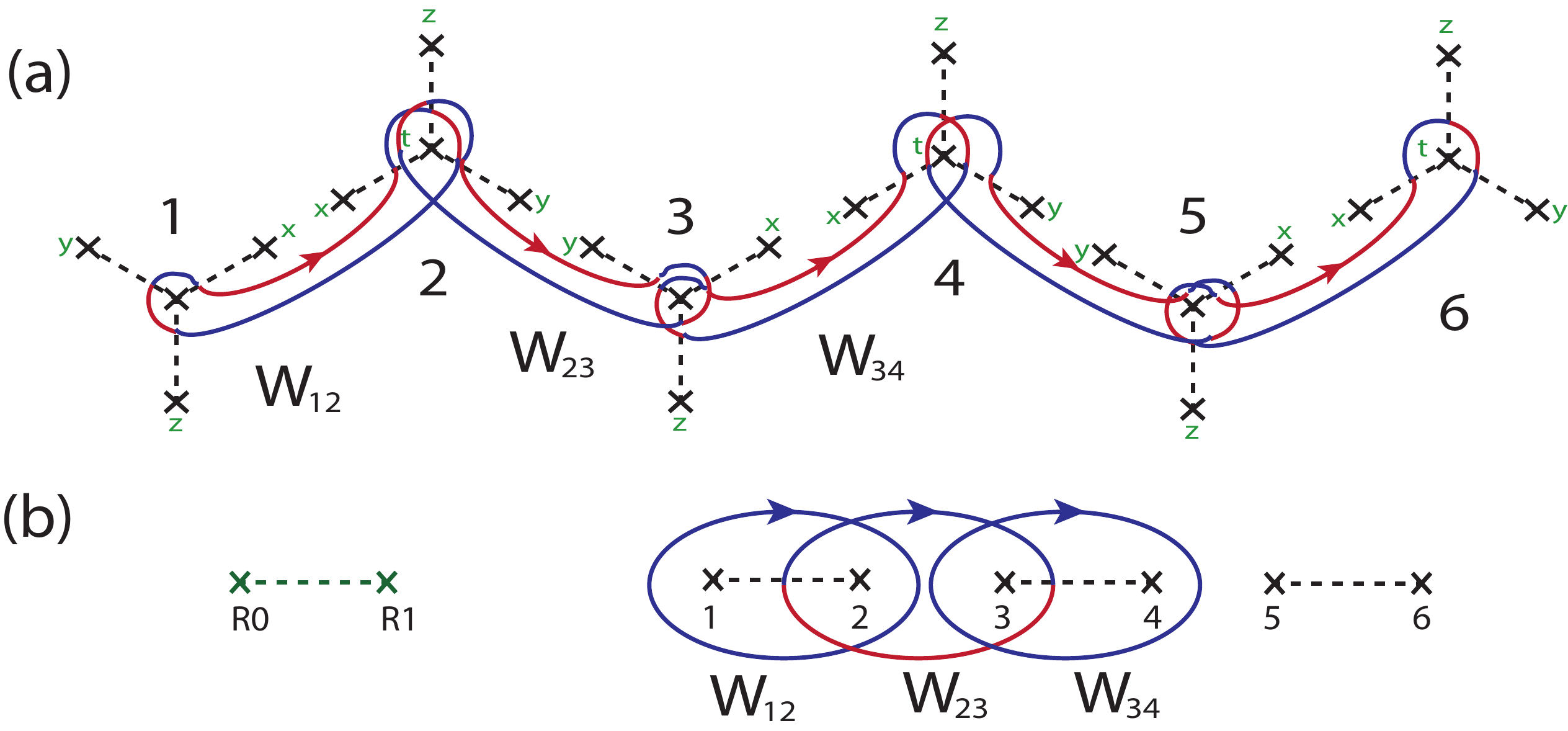}
\caption{
\label{1DchainA} 
(a) The slave genon representation for a single chain of the honeycomb model. The loops
corresponding to the quasiparticle loop operators $W_{ij}$ are shown. 
(b) The Hamiltonian (\ref{1DH}), which describes a 1D Potts model, 
can be reformulated in a slightly different genon representation. 
Each pair of genons, which gives rise to $n$ states, is effectively
a single site of the Potts model. The Wilson loop operators $W_{2i-1,2i}$ and $W_{2i,2i+1}$ are 
shown, and acquire non-trivial commutation relations due to the single crossing. 
A set of reference defects, labelled $R_0$ and $R_1$ and in green color, 
are used as well, although they are not directly associated with any site of the Potts chain. 
They are useful for regulating the strings of the parafermion operators used later. 
Their necessity can be understood by recalling that if we start with $N$ pairs of 
genons on a sphere, the resulting topological degeneracy is $n^{N-1}$ \cite{barkeshli2013genon}. 
Therefore, if the Potts chain has $N$ sites, we need $N+1$ pairs of genons. $R_0$ and $R_1$ 
can be thought of as this extra pair.
}
\end{figure}

\section{Mapping to parafermions and modified Hamiltonian}

\subsection{Mapping to parafermion array}

Let us begin with the model, eq. (1) of the main text, defined along a
one-dimensional chain. As explained in the main text, after applying the 
slave genon transformation and considering the case where the $Z_n$ gauge 
fields are uniformly equal to one, we obtain:
\begin{align}
\label{1DH}
H_{1D} = -\sum_i (J_x W_{2i-1,2i} + J_y W_{2i,2i+1} + H.c.) ,
\end{align}
with 
\begin{align}
W_{2i-1,2i} W_{2i,2i+1} = W_{2i,2i+1} W_{2i-1,2i} \omega. 
\end{align}
It will be useful to introduce a second representation of (\ref{1DH}) in terms of
genons, as shown in Fig. \ref{1DchainA}b. To understand this, suppose that
(\ref{1DH}) contains $2N$ sites of the honeycomb lattice. 
In the second representation, we introduce $N+1$ genons, 
as shown in Fig. \ref{1DchainA}, with the loops $W_{i,i+1}$ as shown. 

$H_{1D}$ is equivalent to the transverse field $Z_n$ Potts model. To see this,
we group the genons $2i-1, 2i$ into a single site with $n$ states, and define
\begin{align}
\tau_i \equiv W_{2i-1,2i}, \;\; \sigma_i^\dagger \sigma_{i+1} \equiv W_{2i,2i+1} ,
\end{align}
such that $\sigma_j \tau_i = \tau_i \sigma_j \omega^{\delta_{ij}}$, 
and $\tau_i^n = \sigma_i^n = 1$, where
$\delta_{ij}$ is the Kronecker delta function. In these variables, 
\begin{align}
H_{1D} = -\sum_i (J_x \tau_i + J_y \sigma_i^\dagger \sigma_{i+1} + H.c.),
\end{align}
which is the familiar form for the $Z_n$ Potts model.  

\begin{figure}
\includegraphics[width=5in]{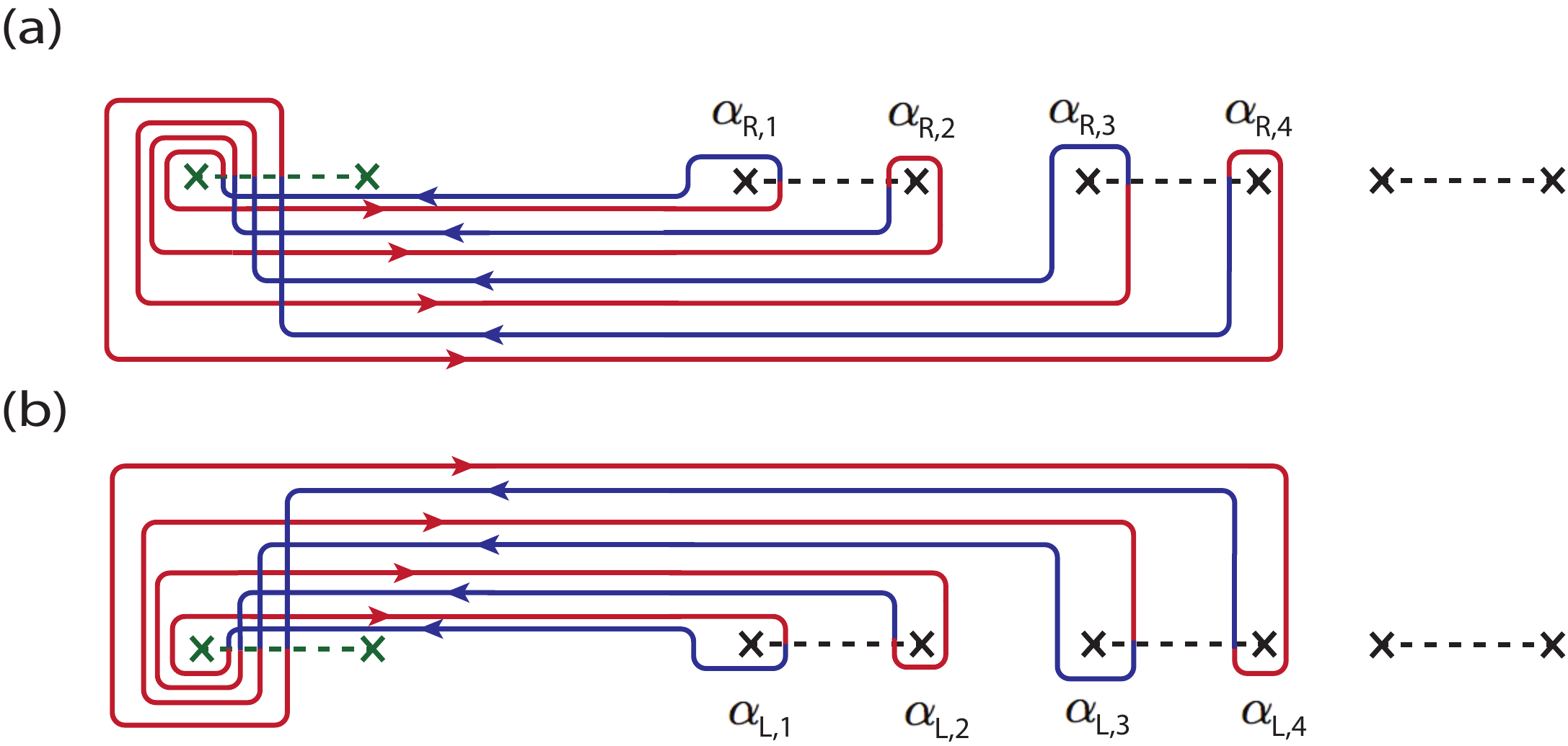}
\caption{
\label{fig:1Dparaf} 
(a) The parafermion operators $\alpha_{R,i}$ correspond to quasiparticle loop operators around the loops shown. 
(b) The parafermion operators $\alpha_{L,i}$ correspond to quasiparticle loop operators around the loops shown. 
}
\end{figure}

We can define ``parafermion'' operators:
\begin{align}
\label{parafDef}
\alpha_{R,2j-1} &= \sigma_j \mu_{j-1}, \;\;
\alpha_{R,2j} = \omega \sigma_j \mu_j,
\nonumber \\
\alpha_{L,2j-1} &= \sigma_j \mu_{j-1}^\dagger, \;\;
\alpha_{L,2j} = \omega^{-1} \sigma_j \mu_j^\dagger,
\end{align}
where $\mu_j \equiv \prod_{k \leq j} \tau_k$. 
The parafermion operators satisfy the algebra:
\begin{align}
\alpha_{Ri} \alpha_{Rj} &= e^{i 2\pi sgn(j-i)/n} \alpha_{Rj} \alpha_{Ri},
\nonumber \\
\alpha_{Li} \alpha_{Lj} &= e^{-i 2\pi sgn(j-i)/n} \alpha_{Lj} \alpha_{Li},
\end{align}
with $(\alpha_{Li})^n = (\alpha_{Ri})^n = 1$. 
Note that $\alpha_L$ and $\alpha_R$ are not independent degrees of freedom. 
In terms of these lattice parafermion operators, 
\begin{align}
\tau_j = \omega^* \alpha_{R,2j-1}^\dagger \alpha_{R,2j} = \omega^* \alpha_{L,2j}^\dagger \alpha_{L,2j-1}
\nonumber \\
\sigma_j^\dagger \sigma_{j+1} = \omega^2 \alpha_{R,2j}^\dagger \alpha_{R,2j+1} = (\omega^*)^2 \alpha_{L,2j}^\dagger \alpha_{L,2j+1}
\end{align}
Therefore, in terms of the lattice parafermions, the Hamiltonian is
\begin{align}
H_{1D} &= -\sum_i (J_x \omega \alpha_{R,2j}^\dagger \alpha_{R,2j-1} + J_y \omega^2 \alpha_{R,2j}^\dagger \alpha_{R,2j+1} + H.c.)
\nonumber \\
&= - \sum_i (J_x \omega^* \alpha_{L,2j}^\dagger \alpha_{L,2j-1} + J_y \omega^{*2} \alpha_{L,2j}^\dagger \alpha_{L,2j+1} + H.c.). 
\end{align}
In Fig. \ref{fig:1Dparaf}, we show how the parafermion operators can be understood in terms of the genon representation as Wilson
loop operators of Abelian quasiparticles. 

Now let us turn to the 2D version of the model, eq. (4) in the main text. Again,
for simplicity we will consider the ground state sector where $u_{ij}$ are uniform in space. 
The 2D model can be understand as an array of 1D chains, together with 
an appropriate interchain coupling:
\begin{align}
H = \sum_m H_{1D}[m] + H_{inter},
\end{align}
where now
\begin{align}
H_{1D}[m] &= -\sum_i (J_x \omega \alpha_{R,2j,m}^\dagger \alpha_{R,2j-1,m} + J_y \omega^2 \alpha_{R,2j,m}^\dagger \alpha_{R,2j+1,m} + H.c.),
\end{align}
and $m$ is the chain index, and
\begin{align}
H_{inter} = -\sum_{\langle i j \rangle = z-link } J_z W_{ij} + H.c.,
\end{align}
where $\langle i j \rangle$ is a vertical $z$-link of the honeycomb lattice. 

\begin{figure}
\includegraphics[width=5in]{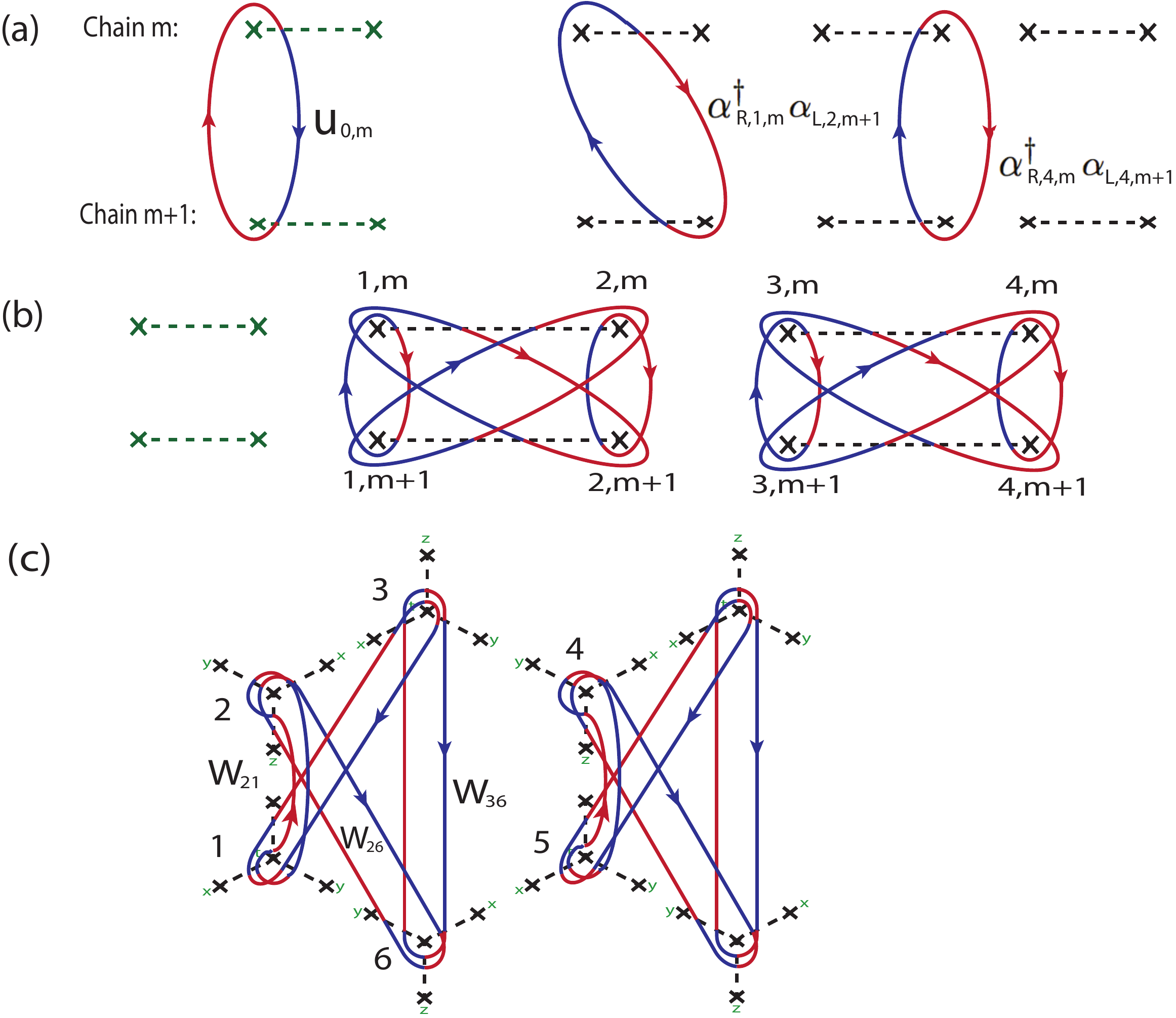}
\caption{
\label{fig:interchain} Depiction of interchain couplings, between chains $m$ and $m+1$. 
(a) The loops corresponding to the operators $\alpha_{R,1,m}^\dagger \alpha_{L,2,m+1}$ and 
$\alpha_{R,4,m}^\dagger \alpha_{L,4,m+1}$ are shown. These loops also contain a loop that encloses
the reference genons from the two chains, which we have labelled $u_{0,m}$. $u_{0,m}$ acts completely 
trivially in the Hilbert space of the Potts chain. 
(b) The loops associated with the interchain parafermion interactions in (\ref{HinterParaf}). 
(c) The equivalent loops of (b), but shown in the original slave genon representation of the honeycomb model. 
We label the associated loop operators as $W_{21}$, $W_{36}$, $W_{26}$, and $W_{31}$. 
}
\end{figure}
From Fig. \ref{fig:interchain}a, it is straightforward to see that the Wilson loop operators which couple
different parafermion chains can be written as
\begin{align}
H_{inter} = -\sum_m J_z (u_{0,m}^\dagger \alpha_{R,2j,m}^\dagger \alpha_{L,2j,m+1} + H.c.),
\end{align}
where $u_{0,m}$ is the loop operator shown in Fig. \ref{fig:interchain}a, which encloses the reference defects. 
Since this loop encloses the reference defects and commutes with all terms in the Hamiltonian of the 
Potts chain, it can be treated as a c-number. 

The parafermion operators $\alpha_{R/L, i}$ can also be written in terms of the original spins 
of the generalized Kitaev model. To understand this, let us first observe that for a single
chain of the generalized Kitaev model (see Fig. 4a), we have the relation:
\begin{align}
\label{T1yW12}
T_1^y W_{12} = W_{12} T_1^y \omega^*. 
\end{align}
In the alternate representation of Fig. 4b, we can therefore associate 
\begin{align}
\alpha_{R1} \propto T_1^{y \dagger}, \;\;
\alpha_{L1} \propto T_1^{y\dagger},
\end{align}
in order to reproduce (\ref{T1yW12}). Furthermore, from
(\ref{parafDef}), we see that in the Potts model $\alpha_{R,1} = \alpha_{L1} = \sigma_1$,
and so
\begin{align}
\alpha_{R,2j-1} &= \alpha_{R1} \prod_{k = 1}^{j-1} \sigma_k^\dagger \sigma_{k+1} \prod_{k =1}^{j-1} \tau_k
\propto T_1^{y\dagger} W_{12} W_{23} ... W_{2j-2,2j-1}
\nonumber \\
\alpha_{R,2j} &= \omega \alpha_{R1} \prod_{k = 1}^{j-1} \sigma_k^\dagger \sigma_{k+1} \prod_{k =1}^{j} \tau_k
\propto T_1^{y\dagger} W_{12} W_{23} ... W_{2j-1,2j}
\nonumber \\
\alpha_{L,2j-1} &= \alpha_{L1} \prod_{k = 1}^{j-1} \sigma_k^\dagger \sigma_{k+1} \prod_{k =1}^{j-1} \tau_k^\dagger
\propto T_1^{y\dagger} W_{12}^\dagger W_{23} W_{34}^\dagger ... W_{2j-2,2j-1}
\nonumber\\
\alpha_{L,2j} &= \omega^* \alpha_{L1} \prod_{k = 1}^{j-1} \sigma_k^\dagger \sigma_{k+1} \prod_{k =1}^{j} \tau_k^\dagger
\propto T_1^{y\dagger} W_{12}^\dagger W_{23} ... W_{2j-1,2j}^\dagger
\end{align}
By including the $Z_3$ gauge fields $u_{ij}$ along the chain, these operators can be made to be gauge-invariant
and therefore expressible in terms of the local spin operators of the generalized Kitaev chain:
\begin{align}
\alpha_{R,j} &\propto T_1^{y\dagger} K_{12} K_{23} ... K_{j-1,j}
\nonumber \\
\alpha_{L,j} &\propto T_1^{y\dagger} K_{12}^\dagger K_{23} K_{34}^\dagger ... K_{j-1,j}^{s_j},
\end{align}
where $s_j = -(-1)^j$. 

\subsection{Controlled limit}

Let us now return to the 1D Hamiltonian, and specialize to the case $n = 3$. 
When $J_x = J_y$, the model is self-dual and lies at a critical point between 
the ordered and disordered phase of the $Z_3$ Potts model. This critical
point is described by a $Z_3$ parafermion conformal field theory. 

As was pointed out recently \cite{mong2013,vaezi2013c}, at the critical point
the lattice parafermion operators can be expanded in terms of the
continuum fields of the $Z_3$ parafermion CFT as:
\begin{align}
\alpha_{R,j} &\sim a \psi_R + (-1)^j b \sigma_R \epsilon_L + ...,
\nonumber \\
\alpha_{L,j} & \sim a \psi_L + (-1)^j b \sigma_L \epsilon_R + ...,
\end{align}
where $\psi_{R/L}$ are the right/left moving $Z_3$ parafermion fields,
$\sigma_{R/L}$ are the $Z_3$ order parameter fields, and $\epsilon_{R/L}$ 
are the energy operators for the right/left moving sectors of the theory. 
$a$ and $b$ are constants. The $...$ include less relevant terms with 
higher scaling dimensions. 

Using the above expansion, let us consider the following interchain
coupling between the uncoupled 1D chains:
\begin{align}
\label{HinterParaf}
H_{inter} &= -\lambda \sum_{j,m} u_{0,m}^\dagger (\alpha_{R,2j,m} + \alpha_{R,2j+1,m})^\dagger (\alpha_{L,2j,m+1} + \alpha_{L,2j+1,m+1}) + H.c.
\nonumber \\
&\sim -4 a^2 \lambda \sum_m u_{0,m}^\dagger (\psi_{R,m}^\dagger \psi_{L,m+1} + H.c.) + ....,
\end{align}
where recall $u_{0,m}^\dagger$ is a c-number here, which we can set to 1. 
When $\lambda > 0$, the above perturbation gaps out counterpropagating 
$Z_3$ parafermion modes from different chains \cite{mong2013}. 
This leaves a gapped two-dimensional bulk, with a chiral $Z_3$ parafermion
mode propagating along the boundary. 

$H_{inter}$ above is written in terms of the lattice parafermion operators. Now we would like
to find the appropriate interchain coupling in terms of the original spin Hamiltonian, which
reduces to $H_{inter}$ after the slave genon transformation for the uniform choice of $Z_3$ 
gauge fields. In order to do this, we use the graphical loop representation developed in
this paper. $H_{inter}$ can be written as
\begin{align}
H_{inter} &= -\lambda \sum_{\hexagon} (W_{21} + W_{31} + W_{26}  + W_{36} + H.c.),
\end{align}
where the loop operators $W_{21}$, $W_{31}$, $W_{61}$, and $W_{36}$ are shown in 
Fig. \ref{fig:interchain}c. 

\begin{figure}
\includegraphics[width=5in]{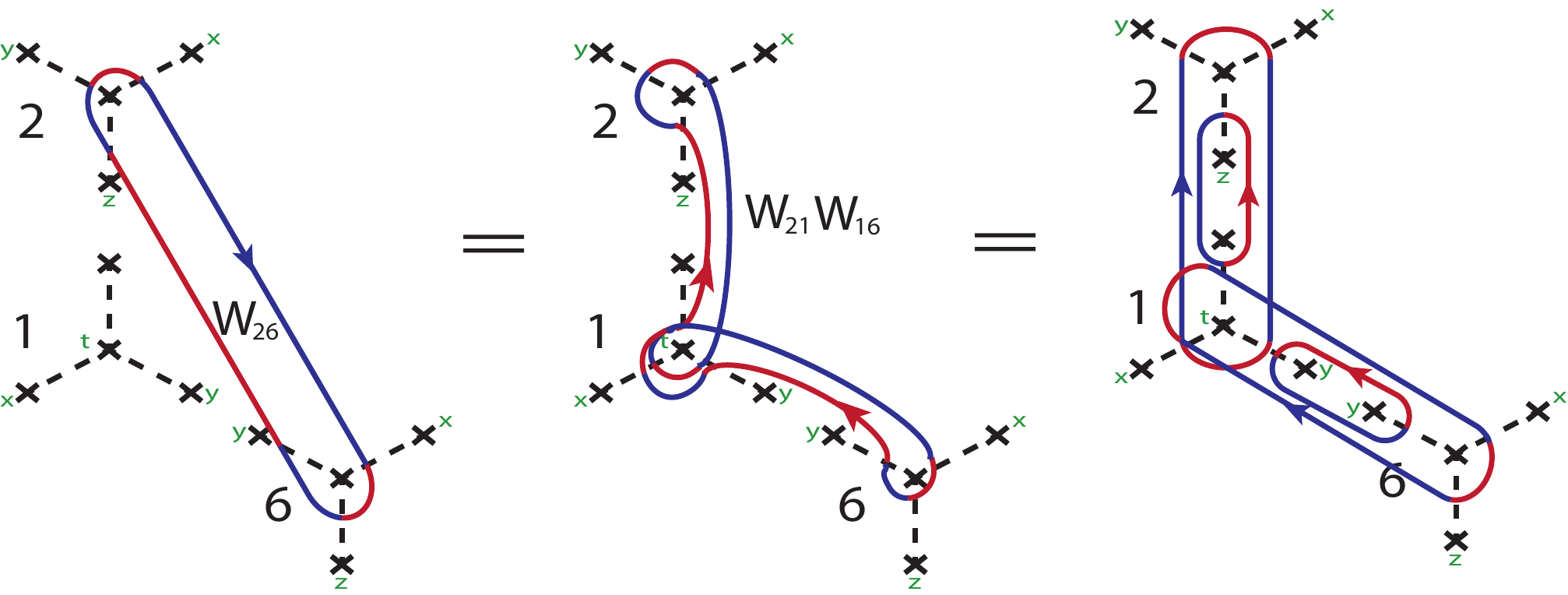}
\caption{
\label{nnnFig} Illustration of the equivalence between $W_{26}$ and 
$u_{21}^\dagger u_{16}^\dagger T_2^z T_1^z T_1^y T_6^y$, allowing us to write down
generic interchain couplings in terms of local spin interactions. 
}
\end{figure}

It is then straightfoward to show (for example, see Fig. \ref{nnnFig}) that the 
following interactions in the spin model:
\begin{align}
\label{HinterSpin}
H_{inter} = -J_z (T^z_1 T^z_2 + T^z_1 T^y_1 T^z_2  T^y_6
+ T^x_3 T^x_2 T^z_2 T^z_1 T^y_1 T^y_6 + T^x_3 T^x_2 T^z_2 T^z_1  + H.c.),
\end{align}
can be rewritten as
\begin{align}
H_{inter} = -J_z (u_{21} W_{21} + u_{21} u_{16} W_{26} + u_{32}u_{21}u_{16} W_{36}  + H.c.) .
\end{align}
Therefore, (\ref{HinterSpin}) reduces to (\ref{HinterParaf}), in the sector with spatially uniform
$Z_3$ gauge fields $u_{ij}$. 

It follows that the modified Hamiltonian,
\begin{align}
H' = -\sum_{\langle i j \rangle} J_{s_{ij}} T^{s_{ij}}_i T^{s_{ij}}_j + H.c.
- J_z \sum_{\hexagon} \mathcal{O}_{\hexagon},
\end{align}
with
\begin{align}
\mathcal{O}_{\hexagon} = T^z_1 T^y_1 T^z_2  T^y_6
+ T^x_3 T^x_2 T^z_2 T^z_1 T^y_1 T^y_6 + T^x_3 T^x_2 T^z_2 T^z_1  + H.c.,
\end{align}
is expected to realize a non-Abelian topologically ordered phase when
$J_z \ll J_x = J_y$, and $J_z, J_x, J_y > 0$. This non-Abelian phase has a 
chiral $Z_3$ parafermion CFT propagating along its boundary. Therefore, 
it contains a non-Abelian ``Fibonacci'' anyon. 

%\bibliography{TI}

\begin{thebibliography}{49}
\expandafter\ifx\csname natexlab\endcsname\relax\def\natexlab#1{#1}\fi
\expandafter\ifx\csname bibnamefont\endcsname\relax
  \def\bibnamefont#1{#1}\fi
\expandafter\ifx\csname bibfnamefont\endcsname\relax
  \def\bibfnamefont#1{#1}\fi
\expandafter\ifx\csname citenamefont\endcsname\relax
  \def\citenamefont#1{#1}\fi
\expandafter\ifx\csname url\endcsname\relax
  \def\url#1{\texttt{#1}}\fi
\expandafter\ifx\csname urlprefix\endcsname\relax\def\urlprefix{URL }\fi
\providecommand{\bibinfo}[2]{#2}
\providecommand{\eprint}[2][]{\url{#2}}

\bibitem[{\citenamefont{Kitaev}(2006)}]{kitaev2006}
\bibinfo{author}{\bibfnamefont{A.}~\bibnamefont{Kitaev}},
  \bibinfo{journal}{Annals of Physics} \textbf{\bibinfo{volume}{321}},
  \bibinfo{pages}{2 } (\bibinfo{year}{2006}).

\bibitem[{\citenamefont{Nussinov and van~den Brink}(2013)}]{nussinov2013}
\bibinfo{author}{\bibfnamefont{Z.}~\bibnamefont{Nussinov}} \bibnamefont{and}
  \bibinfo{author}{\bibfnamefont{J.}~\bibnamefont{van~den Brink}}
  (\bibinfo{year}{2013}), \eprint{arXiv:1303.5922}.

\bibitem[{\citenamefont{Chaloupka et~al.}(2010)\citenamefont{Chaloupka,
  Jackeli, and Khaliullin}}]{chaloupka2010}
\bibinfo{author}{\bibfnamefont{J.}~\bibnamefont{Chaloupka}},
  \bibinfo{author}{\bibfnamefont{G.}~\bibnamefont{Jackeli}}, \bibnamefont{and}
  \bibinfo{author}{\bibfnamefont{G.}~\bibnamefont{Khaliullin}},
  \bibinfo{journal}{Phys. Rev. Lett.} \textbf{\bibinfo{volume}{105}},
  \bibinfo{pages}{027204} (\bibinfo{year}{2010}).

\bibitem[{\citenamefont{Singh et~al.}(2012)\citenamefont{Singh, Manni, Reuther,
  Berlijn, Thomale, Ku, Trebst, and Gegenwart}}]{singh2012}
\bibinfo{author}{\bibfnamefont{Y.}~\bibnamefont{Singh}},
  \bibinfo{author}{\bibfnamefont{S.}~\bibnamefont{Manni}},
  \bibinfo{author}{\bibfnamefont{J.}~\bibnamefont{Reuther}},
  \bibinfo{author}{\bibfnamefont{T.}~\bibnamefont{Berlijn}},
  \bibinfo{author}{\bibfnamefont{R.}~\bibnamefont{Thomale}},
  \bibinfo{author}{\bibfnamefont{W.}~\bibnamefont{Ku}},
  \bibinfo{author}{\bibfnamefont{S.}~\bibnamefont{Trebst}}, \bibnamefont{and}
  \bibinfo{author}{\bibfnamefont{P.}~\bibnamefont{Gegenwart}},
  \bibinfo{journal}{Phys. Rev. Lett.} \textbf{\bibinfo{volume}{108}},
  \bibinfo{pages}{127203} (\bibinfo{year}{2012}).

\bibitem[{\citenamefont{Duan et~al.}(2003)\citenamefont{Duan, Demler, and
  Lukin}}]{duan2003}
\bibinfo{author}{\bibfnamefont{L.-M.} \bibnamefont{Duan}},
  \bibinfo{author}{\bibfnamefont{E.}~\bibnamefont{Demler}}, \bibnamefont{and}
  \bibinfo{author}{\bibfnamefont{M.~D.} \bibnamefont{Lukin}},
  \bibinfo{journal}{Phys. Rev. Lett.} \textbf{\bibinfo{volume}{91}},
  \bibinfo{pages}{090402} (\bibinfo{year}{2003}).

\bibitem[{\citenamefont{Nayak et~al.}(2008)\citenamefont{Nayak, Simon, Stern,
  Freedman, and Sarma}}]{nayak2008}
\bibinfo{author}{\bibfnamefont{C.}~\bibnamefont{Nayak}},
  \bibinfo{author}{\bibfnamefont{S.~H.} \bibnamefont{Simon}},
  \bibinfo{author}{\bibfnamefont{A.}~\bibnamefont{Stern}},
  \bibinfo{author}{\bibfnamefont{M.}~\bibnamefont{Freedman}}, \bibnamefont{and}
  \bibinfo{author}{\bibfnamefont{S.~D.} \bibnamefont{Sarma}},
  \bibinfo{journal}{Rev. Mod. Phys.} \textbf{\bibinfo{volume}{80}},
  \bibinfo{pages}{1083} (\bibinfo{year}{2008}).

\bibitem[{\citenamefont{Barkeshli and Qi}(2012)}]{barkeshli2012a}
\bibinfo{author}{\bibfnamefont{M.}~\bibnamefont{Barkeshli}} \bibnamefont{and}
  \bibinfo{author}{\bibfnamefont{X.-L.} \bibnamefont{Qi}},
  \bibinfo{journal}{Phys. Rev. X} \textbf{\bibinfo{volume}{2}},
  \bibinfo{pages}{031013} (\bibinfo{year}{2012}), \eprint{arXiv:1112.3311}.

\bibitem[{\citenamefont{Barkeshli
  et~al.}(2013{\natexlab{a}})\citenamefont{Barkeshli, Jian, and
  Qi}}]{barkeshli2013genon}
\bibinfo{author}{\bibfnamefont{M.}~\bibnamefont{Barkeshli}},
  \bibinfo{author}{\bibfnamefont{C.-M.} \bibnamefont{Jian}}, \bibnamefont{and}
  \bibinfo{author}{\bibfnamefont{X.-L.} \bibnamefont{Qi}},
  \bibinfo{journal}{Phys. Rev. B} \textbf{\bibinfo{volume}{87}},
  \bibinfo{pages}{045130} (\bibinfo{year}{2013}{\natexlab{a}}),
  \eprint{arXiv:1208.4834}.

\bibitem[{\citenamefont{Fendley}(2012)}]{fendley2012}
\bibinfo{author}{\bibfnamefont{P.}~\bibnamefont{Fendley}}, \bibinfo{journal}{J.
  Stat. Mech.} p. \bibinfo{pages}{P11020} (\bibinfo{year}{2012}).

\bibitem[{\citenamefont{Lindner et~al.}(2012)\citenamefont{Lindner, Berg,
  Refael, and Stern}}]{lindner2012}
\bibinfo{author}{\bibfnamefont{N.~H.} \bibnamefont{Lindner}},
  \bibinfo{author}{\bibfnamefont{E.}~\bibnamefont{Berg}},
  \bibinfo{author}{\bibfnamefont{G.}~\bibnamefont{Refael}}, \bibnamefont{and}
  \bibinfo{author}{\bibfnamefont{A.}~\bibnamefont{Stern}},
  \bibinfo{journal}{Phys. Rev. X} \textbf{\bibinfo{volume}{2}},
  \bibinfo{pages}{041002} (\bibinfo{year}{2012}), \eprint{arXiv:1204.5733}.

\bibitem[{\citenamefont{Clarke et~al.}(2013)\citenamefont{Clarke, Alicea, and
  Shtengel}}]{clarke2013}
\bibinfo{author}{\bibfnamefont{D.~J.} \bibnamefont{Clarke}},
  \bibinfo{author}{\bibfnamefont{J.}~\bibnamefont{Alicea}}, \bibnamefont{and}
  \bibinfo{author}{\bibfnamefont{K.}~\bibnamefont{Shtengel}},
  \bibinfo{journal}{Nature Comm.} \textbf{\bibinfo{volume}{4}},
  \bibinfo{pages}{1348} (\bibinfo{year}{2013}), \eprint{arXiv:1204.5479}.

\bibitem[{\citenamefont{Cheng}(2012)}]{cheng2012}
\bibinfo{author}{\bibfnamefont{M.}~\bibnamefont{Cheng}},
  \bibinfo{journal}{Phys. Rev. B} \textbf{\bibinfo{volume}{86}},
  \bibinfo{pages}{195126} (\bibinfo{year}{2012}), \eprint{arXiv:1204.6084}.

\bibitem[{\citenamefont{Vaezi}(2013{\natexlab{a}})}]{vaezi2013}
\bibinfo{author}{\bibfnamefont{A.}~\bibnamefont{Vaezi}},
  \bibinfo{journal}{Phys. Rev. B} \textbf{\bibinfo{volume}{87}},
  \bibinfo{pages}{035132} (\bibinfo{year}{2013}{\natexlab{a}}).

\bibitem[{\citenamefont{Barkeshli
  et~al.}(2013{\natexlab{b}})\citenamefont{Barkeshli, Jian, and
  Qi}}]{barkeshli2013defect}
\bibinfo{author}{\bibfnamefont{M.}~\bibnamefont{Barkeshli}},
  \bibinfo{author}{\bibfnamefont{C.-M.} \bibnamefont{Jian}}, \bibnamefont{and}
  \bibinfo{author}{\bibfnamefont{X.-L.} \bibnamefont{Qi}},
  \bibinfo{journal}{Phys. Rev. B} \textbf{\bibinfo{volume}{88}},
  \bibinfo{pages}{241103(R)} (\bibinfo{year}{2013}{\natexlab{b}});
\bibinfo{author}{\bibfnamefont{M.}~\bibnamefont{Barkeshli}},
  \bibinfo{author}{\bibfnamefont{C.-M.} \bibnamefont{Jian}}, \bibnamefont{and}
  \bibinfo{author}{\bibfnamefont{X.-L.} \bibnamefont{Qi}},
  \bibinfo{journal}{Phys. Rev. B} \textbf{\bibinfo{volume}{88}},
  \bibinfo{pages}{235103} (\bibinfo{year}{2013}{\natexlab{c}}).

\bibitem[{\citenamefont{Mong et~al.}()\citenamefont{Mong, Clarke, Alicea,
  Lindner, Fendley, Nayak, Yuval~Oreg, Berg, Shtengel, and Fisher}}]{mong2013}
\bibinfo{author}{\bibfnamefont{R.~S.~K.} \bibnamefont{Mong}},
  \bibinfo{author}{\bibfnamefont{D.~J.} \bibnamefont{Clarke}},
  \bibinfo{author}{\bibfnamefont{J.}~\bibnamefont{Alicea}},
  \bibinfo{author}{\bibfnamefont{N.~H.} \bibnamefont{Lindner}},
  \bibinfo{author}{\bibfnamefont{P.}~\bibnamefont{Fendley}},
  \bibinfo{author}{\bibfnamefont{C.}~\bibnamefont{Nayak}},
  \bibinfo{author}{\bibfnamefont{A.~S.} \bibnamefont{Yuval~Oreg}},
  \bibinfo{author}{\bibfnamefont{E.}~\bibnamefont{Berg}},
  \bibinfo{author}{\bibfnamefont{K.}~\bibnamefont{Shtengel}}, \bibnamefont{and}
  \bibinfo{author}{\bibfnamefont{M.~P.~A.} \bibnamefont{Fisher}},
  \bibinfo{note}{arXiv:1307.4403}.

\bibitem[{sup()}]{supplement}
\bibinfo{note}{See the supplementary material for further information.}

\bibitem[{\citenamefont{Kitaev}(2003)}]{kitaev2003}
\bibinfo{author}{\bibfnamefont{A.}~\bibnamefont{Kitaev}},
  \bibinfo{journal}{Annals Phys.} \textbf{\bibinfo{volume}{303}},
  \bibinfo{pages}{2} (\bibinfo{year}{2003}), \eprint{arXiv:quant-ph/9707021}.

\bibitem[{\citenamefont{Wen}(2003)}]{wen2003}
\bibinfo{author}{\bibfnamefont{X.-G.} \bibnamefont{Wen}},
  \bibinfo{journal}{Phys. Rev. Lett.} \textbf{\bibinfo{volume}{90}},
  \bibinfo{pages}{016803} (\bibinfo{year}{2003}).

\bibitem[{\citenamefont{Schulz et~al.}(2012)\citenamefont{Schulz, Dusuel, Orus,
  Vidal, and Schmidt}}]{schulz2012}
\bibinfo{author}{\bibfnamefont{M.~D.} \bibnamefont{Schulz}},
  \bibinfo{author}{\bibfnamefont{S.}~\bibnamefont{Dusuel}},
  \bibinfo{author}{\bibfnamefont{R.}~\bibnamefont{Orus}},
  \bibinfo{author}{\bibfnamefont{J.}~\bibnamefont{Vidal}}, \bibnamefont{and}
  \bibinfo{author}{\bibfnamefont{K.~P.} \bibnamefont{Schmidt}},
  \bibinfo{journal}{New Journal of Physics} \textbf{\bibinfo{volume}{14}},
  \bibinfo{pages}{025005} (\bibinfo{year}{2012}).

\bibitem[{\citenamefont{Barkeshli and Wen}(2010)}]{barkeshli2010}
\bibinfo{author}{\bibfnamefont{M.}~\bibnamefont{Barkeshli}} \bibnamefont{and}
  \bibinfo{author}{\bibfnamefont{X.-G.} \bibnamefont{Wen}},
  \bibinfo{journal}{Phys. Rev. B} \textbf{\bibinfo{volume}{81}},
  \bibinfo{pages}{045323} (\bibinfo{year}{2010}), \eprint{arXiv:0909.4882}.

\bibitem[{\citenamefont{Bombin}(2010)}]{bombin2010}
\bibinfo{author}{\bibfnamefont{H.}~\bibnamefont{Bombin}},
  \bibinfo{journal}{Phys. Rev. Lett.} \textbf{\bibinfo{volume}{105}},
  \bibinfo{pages}{030403} (\bibinfo{year}{2010}), \eprint{arXiv:1004.1838}.

\bibitem[{\citenamefont{Kitaev and Kong}(2012)}]{kitaev2012}
\bibinfo{author}{\bibfnamefont{A.}~\bibnamefont{Kitaev}} \bibnamefont{and}
  \bibinfo{author}{\bibfnamefont{L.}~\bibnamefont{Kong}},
  \bibinfo{journal}{Comm. Math. Phys.} \textbf{\bibinfo{volume}{313}},
  \bibinfo{pages}{351} (\bibinfo{year}{2012}).

\bibitem[{\citenamefont{You and Wen}(2012)}]{you2012}
\bibinfo{author}{\bibfnamefont{Y.-Z.} \bibnamefont{You}} \bibnamefont{and}
  \bibinfo{author}{\bibfnamefont{X.-G.} \bibnamefont{Wen}},
  \bibinfo{journal}{Phys. Rev. B} \textbf{\bibinfo{volume}{86}},
  \bibinfo{pages}{161107} (\bibinfo{year}{2012}).

\bibitem[{\citenamefont{Teo et~al.}(2013)\citenamefont{Teo, Roy, and
  Chen}}]{teo2013}
\bibinfo{author}{\bibfnamefont{J.~C.} \bibnamefont{Teo}},
  \bibinfo{author}{\bibfnamefont{A.}~\bibnamefont{Roy}}, \bibnamefont{and}
  \bibinfo{author}{\bibfnamefont{X.}~\bibnamefont{Chen}}
  (\bibinfo{year}{2013}), \eprint{arXiv:1308.5984}.

\bibitem[{\citenamefont{Brown et~al.}(2013)\citenamefont{Brown, Bartlett,
  Doherty, and Barrett}}]{brown2013}
\bibinfo{author}{\bibfnamefont{B.~J.} \bibnamefont{Brown}},
  \bibinfo{author}{\bibfnamefont{S.~D.} \bibnamefont{Bartlett}},
  \bibinfo{author}{\bibfnamefont{A.~C.} \bibnamefont{Doherty}},
  \bibnamefont{and} \bibinfo{author}{\bibfnamefont{S.~D.}
  \bibnamefont{Barrett}} (\bibinfo{year}{2013}), \eprint{arXiv:1303.4455}.

\bibitem[{\citenamefont{Barkeshli and Qi}(2013)}]{barkeshli2013}
\bibinfo{author}{\bibfnamefont{M.}~\bibnamefont{Barkeshli}} \bibnamefont{and}
  \bibinfo{author}{\bibfnamefont{X.-L.} \bibnamefont{Qi}}
  (\bibinfo{year}{2013}), \eprint{arXiv:1302.2673}.

\bibitem[{\citenamefont{Khan et~al.}(2014)\citenamefont{Khan, Teo, and
  Hughes}}]{khan2014}
\bibinfo{author}{\bibfnamefont{M.~N.} \bibnamefont{Khan}},
  \bibinfo{author}{\bibfnamefont{J.~C.~Y.} \bibnamefont{Teo}},
  \bibnamefont{and} \bibinfo{author}{\bibfnamefont{T.~L.} \bibnamefont{Hughes}}
  (\bibinfo{year}{2014}), \eprint{arXiv:1403.6478}.

\bibitem[{\citenamefont{Wen}(2004)}]{wen04}
\bibinfo{author}{\bibfnamefont{X.-G.} \bibnamefont{Wen}},
  \emph{\bibinfo{title}{Quantum Field Theory of Many-Body Systems}}
  (\bibinfo{publisher}{Oxford Univ. Press}, \bibinfo{address}{Oxford},
  \bibinfo{year}{2004}).

\bibitem[{\citenamefont{Fradkin and Kadanoff}(1980)}]{fradkin1980}
\bibinfo{author}{\bibfnamefont{E.}~\bibnamefont{Fradkin}} \bibnamefont{and}
  \bibinfo{author}{\bibfnamefont{L.~P.} \bibnamefont{Kadanoff}},
  \bibinfo{journal}{Nucl. Phys. B} \textbf{\bibinfo{volume}{170}},
  \bibinfo{pages}{1} (\bibinfo{year}{1980}).

\bibitem[{\citenamefont{Fendley}(2014)}]{fendley2014}
\bibinfo{author}{\bibfnamefont{P.}~\bibnamefont{Fendley}},
  \bibinfo{journal}{Journal of Physics A: Mathematical and Theoretical}
  \textbf{\bibinfo{volume}{47}}, \bibinfo{pages}{075001}
  (\bibinfo{year}{2014}).

\bibitem[{\citenamefont{Burrello et~al.}(2013)\citenamefont{Burrello, van Heck,
  and Cobanera}}]{burrello2013}
\bibinfo{author}{\bibfnamefont{M.}~\bibnamefont{Burrello}},
  \bibinfo{author}{\bibfnamefont{B.}~\bibnamefont{van Heck}}, \bibnamefont{and}
  \bibinfo{author}{\bibfnamefont{E.}~\bibnamefont{Cobanera}},
  \bibinfo{journal}{Phys. Rev. B} \textbf{\bibinfo{volume}{87}},
  \bibinfo{pages}{195422} (\bibinfo{year}{2013}).

\bibitem[{\citenamefont{Bondesan and Quella}(2013)}]{bondesan2013}
\bibinfo{author}{\bibfnamefont{R.}~\bibnamefont{Bondesan}} \bibnamefont{and}
  \bibinfo{author}{\bibfnamefont{T.}~\bibnamefont{Quella}},
  \bibinfo{journal}{Journal of Statistical Mechanics: Theory and Experiment}
  \textbf{\bibinfo{volume}{2013}}, \bibinfo{pages}{P10024}
  (\bibinfo{year}{2013}).

\bibitem[{\citenamefont{Vaezi}(2013{\natexlab{b}})}]{vaezi2013b}
\bibinfo{author}{\bibfnamefont{A.}~\bibnamefont{Vaezi}}
  (\bibinfo{year}{2013}{\natexlab{b}}), \eprint{arXiv:1307.8069}.

\bibitem[{\citenamefont{Zamolodchikov and Fateev}(1985)}]{zamolodchikov1985}
\bibinfo{author}{\bibfnamefont{A.}~\bibnamefont{Zamolodchikov}}
  \bibnamefont{and} \bibinfo{author}{\bibfnamefont{V.}~\bibnamefont{Fateev}},
  \bibinfo{journal}{Sov. Phys. JETP} \textbf{\bibinfo{volume}{62}},
  \bibinfo{pages}{215} (\bibinfo{year}{1985}).

\bibitem[{\citenamefont{Yakovenko}(1991)}]{yakovenko1991}
\bibinfo{author}{\bibfnamefont{V.~M.} \bibnamefont{Yakovenko}},
  \bibinfo{journal}{Phys. Rev. B} \textbf{\bibinfo{volume}{43}},
  \bibinfo{pages}{11353} (\bibinfo{year}{1991});
\bibinfo{author}{\bibfnamefont{S.~L.} \bibnamefont{Sondhi}} \bibnamefont{and}
  \bibinfo{author}{\bibfnamefont{K.}~\bibnamefont{Yang}},
  \bibinfo{journal}{Phys. Rev. B} \textbf{\bibinfo{volume}{63}},
  \bibinfo{pages}{054430} (\bibinfo{year}{2001});
\bibinfo{author}{\bibfnamefont{C.~L.} \bibnamefont{Kane}},
  \bibinfo{author}{\bibfnamefont{R.}~\bibnamefont{Mukhopadhyay}},
  \bibnamefont{and} \bibinfo{author}{\bibfnamefont{T.~C.}
  \bibnamefont{Lubensky}}, \bibinfo{journal}{Phys. Rev. Lett.}
  \textbf{\bibinfo{volume}{88}}, \bibinfo{pages}{036401}
  (\bibinfo{year}{2002});
\bibinfo{author}{\bibfnamefont{J.~C.~Y.} \bibnamefont{Teo}} \bibnamefont{and}
  \bibinfo{author}{\bibfnamefont{C.~L.} \bibnamefont{Kane}},
  \bibinfo{journal}{Phys. Rev. B} \textbf{\bibinfo{volume}{89}},
  \bibinfo{pages}{085101} (\bibinfo{year}{2014}), \eprint{arXiv:1111.2617};
\bibinfo{author}{\bibfnamefont{J.}~\bibnamefont{Klinovaja}} \bibnamefont{and}
  \bibinfo{author}{\bibfnamefont{D.}~\bibnamefont{Loss}}
  (\bibinfo{year}{2013}), \eprint{arXiv:1305.1569}.

\bibitem[{\citenamefont{Neupert et~al.}()\citenamefont{Neupert, Chamon, Mudry,
  and Thomale}}]{titus1403}
\bibinfo{author}{\bibfnamefont{T.}~\bibnamefont{Neupert}},
  \bibinfo{author}{\bibfnamefont{C.}~\bibnamefont{Chamon}},
  \bibinfo{author}{\bibfnamefont{C.}~\bibnamefont{Mudry}}, \bibnamefont{and}
  \bibinfo{author}{\bibfnamefont{R.}~\bibnamefont{Thomale}},
  \bibinfo{note}{arXiv:1403.0953}.

\bibitem[{\citenamefont{White}(1992)}]{White1992DMRG}
\bibinfo{author}{\bibfnamefont{S.~R.} \bibnamefont{White}},
  \bibinfo{journal}{Phys. Rev. Lett.} \textbf{\bibinfo{volume}{69}},
  \bibinfo{pages}{2863} (\bibinfo{year}{1992}).

\bibitem[{\citenamefont{Jiang and Balents}()}]{JiangBalents}
\bibinfo{author}{\bibfnamefont{H.~C.} \bibnamefont{Jiang}} \bibnamefont{and}
  \bibinfo{author}{\bibfnamefont{L.}~\bibnamefont{Balents}},
  \bibinfo{note}{arXiv:1309.7438}.

\bibitem[{\citenamefont{{Vaezi} and {Kim}}(2013)}]{vaezi2013c}
\bibinfo{author}{\bibfnamefont{A.}~\bibnamefont{{Vaezi}}} \bibnamefont{and}
  \bibinfo{author}{\bibfnamefont{E.-A.} \bibnamefont{{Kim}}}
  (\bibinfo{year}{2013}), \eprint{arXiv:1310.7434}.

\bibitem[{\citenamefont{Francesco et~al.}(1997)\citenamefont{Francesco,
  Mathieu, and Senechal}}]{difrancesco}
\bibinfo{author}{\bibfnamefont{P.~D.} \bibnamefont{Francesco}},
  \bibinfo{author}{\bibfnamefont{P.}~\bibnamefont{Mathieu}}, \bibnamefont{and}
  \bibinfo{author}{\bibfnamefont{D.}~\bibnamefont{Senechal}},
  \emph{\bibinfo{title}{Conformal Field Theory}}
  (\bibinfo{publisher}{Springer}, \bibinfo{year}{1997}).

\bibitem[{\citenamefont{Feiguin et~al.}(2007)\citenamefont{Feiguin, Trebst,
  Ludwig, Troyer, Kitaev, Wang, and Freedman}}]{adrian2007}
\bibinfo{author}{\bibfnamefont{A.}~\bibnamefont{Feiguin}},
  \bibinfo{author}{\bibfnamefont{S.}~\bibnamefont{Trebst}},
  \bibinfo{author}{\bibfnamefont{A.~W.~W.} \bibnamefont{Ludwig}},
  \bibinfo{author}{\bibfnamefont{M.}~\bibnamefont{Troyer}},
  \bibinfo{author}{\bibfnamefont{A.}~\bibnamefont{Kitaev}},
  \bibinfo{author}{\bibfnamefont{Z.}~\bibnamefont{Wang}}, \bibnamefont{and}
  \bibinfo{author}{\bibfnamefont{M.~H.} \bibnamefont{Freedman}},
  \bibinfo{journal}{Phys. Rev. Lett.} \textbf{\bibinfo{volume}{98}},
  \bibinfo{pages}{160409} (\bibinfo{year}{2007}).

\bibitem[{\citenamefont{Gils et~al.}(2009)\citenamefont{Gils, Trebst, Kitaev,
  Ludwig, Troyer, and Wang}}]{gils2009}
\bibinfo{author}{\bibfnamefont{C.}~\bibnamefont{Gils}},
  \bibinfo{author}{\bibfnamefont{S.}~\bibnamefont{Trebst}},
  \bibinfo{author}{\bibfnamefont{A.}~\bibnamefont{Kitaev}},
  \bibinfo{author}{\bibfnamefont{A.~W.} \bibnamefont{Ludwig}},
  \bibinfo{author}{\bibfnamefont{M.}~\bibnamefont{Troyer}}, \bibnamefont{and}
  \bibinfo{author}{\bibfnamefont{Z.}~\bibnamefont{Wang}},
  \bibinfo{journal}{Nature Physics} \textbf{\bibinfo{volume}{5}},
  \bibinfo{pages}{834} (\bibinfo{year}{2009}).

\bibitem[{sim()}]{simonstalk}
\bibinfo{note}{Xiao-Liang Qi, presentation on the workshop {\it Topological
  Phases of Matter Workshop} at Simons Center, Stony Brook University, June
  13th, 2013. \url{http://scgp.stonybrook.edu/archives/3464}}.

\bibitem[{\citenamefont{Vaezi}()}]{Vaezi}
\bibinfo{author}{\bibfnamefont{A.}~\bibnamefont{Vaezi}},
  \bibinfo{note}{unpublished, submitted to arxiv.}

\end{thebibliography}

\begin{thebibliography}{3}
\expandafter\ifx\csname natexlab\endcsname\relax\def\natexlab#1{#1}\fi
\expandafter\ifx\csname bibnamefont\endcsname\relax
  \def\bibnamefont#1{#1}\fi
\expandafter\ifx\csname bibfnamefont\endcsname\relax
  \def\bibfnamefont#1{#1}\fi
\expandafter\ifx\csname citenamefont\endcsname\relax
  \def\citenamefont#1{#1}\fi
\expandafter\ifx\csname url\endcsname\relax
  \def\url#1{\texttt{#1}}\fi
\expandafter\ifx\csname urlprefix\endcsname\relax\def\urlprefix{URL }\fi
\providecommand{\bibinfo}[2]{#2}
\providecommand{\eprint}[2][]{\url{#2}}

\bibitem[{\citenamefont{Barkeshli et~al.}(2013)\citenamefont{Barkeshli, Jian,
  and Qi}}]{barkeshli2013genon}
\bibinfo{author}{\bibfnamefont{M.}~\bibnamefont{Barkeshli}},
  \bibinfo{author}{\bibfnamefont{C.-M.} \bibnamefont{Jian}}, \bibnamefont{and}
  \bibinfo{author}{\bibfnamefont{X.-L.} \bibnamefont{Qi}},
  \bibinfo{journal}{Phys. Rev. B} \textbf{\bibinfo{volume}{87}},
  \bibinfo{pages}{045130} (\bibinfo{year}{2013}), \eprint{arXiv:1208.4834}.

\bibitem[{\citenamefont{Mong et~al.}()\citenamefont{Mong, Clarke, Alicea,
  Lindner, Fendley, Nayak, Yuval~Oreg, Berg, Shtengel, and Fisher}}]{mong2013}
\bibinfo{author}{\bibfnamefont{R.~S.~K.} \bibnamefont{Mong}},
  \bibinfo{author}{\bibfnamefont{D.~J.} \bibnamefont{Clarke}},
  \bibinfo{author}{\bibfnamefont{J.}~\bibnamefont{Alicea}},
  \bibinfo{author}{\bibfnamefont{N.~H.} \bibnamefont{Lindner}},
  \bibinfo{author}{\bibfnamefont{P.}~\bibnamefont{Fendley}},
  \bibinfo{author}{\bibfnamefont{C.}~\bibnamefont{Nayak}},
  \bibinfo{author}{\bibfnamefont{A.~S.} \bibnamefont{Yuval~Oreg}},
  \bibinfo{author}{\bibfnamefont{E.}~\bibnamefont{Berg}},
  \bibinfo{author}{\bibfnamefont{K.}~\bibnamefont{Shtengel}}, \bibnamefont{and}
  \bibinfo{author}{\bibfnamefont{M.~P.~A.} \bibnamefont{Fisher}},
  \bibinfo{note}{arXiv:1307.4403}.

\bibitem[{\citenamefont{{Vaezi} and {Kim}}(2013)}]{vaezi2013c}
\bibinfo{author}{\bibfnamefont{A.}~\bibnamefont{{Vaezi}}} \bibnamefont{and}
  \bibinfo{author}{\bibfnamefont{E.-A.} \bibnamefont{{Kim}}}
  (\bibinfo{year}{2013}), \eprint{arXiv:1310.7434}.

\end{thebibliography}
\begin{comment}

\end{comment}
\end{widetext}
%\end{document}

\end{document}